\documentclass[aps,pre,reprint,groupedaddress,floatfix,showpacs,showkeys,10pt]{revtex4-1}
\usepackage[utf8]{inputenc}
\usepackage{amsmath}
\usepackage{amssymb}
\usepackage{graphicx}
\usepackage[american]{babel}
\usepackage[amssymb]{SIunits}
\usepackage{subfigure}
\usepackage{notoccite}
\usepackage{afterpage}
\DeclareMathOperator\erfc{erfc}

\newcommand{\vc}{\mbox{\boldmath{$c$}}}
\newcommand{\vmu}{\mbox{\boldmath{$\mu$}}}
\usepackage{tikz}
\usepackage{ulem}
\usetikzlibrary{calc}
\usetikzlibrary{plotmarks}
\usepackage{pgfplots}
\usepackage{etoolbox}
\patchcmd{\normalsize}{13.6}{13}{}{}
\pgfplotsset{
tick label style={font=\tiny},
label style={font=\small},
legend style={font=\tiny}
}

\begin{document}
\title{Growth in multi-component alloys: Theoretical and numerical determination of phase concentrations}
\author{Arka Lahiri, T.A.Abinandanan, Abhik Choudhury, M.S. Bhaskar}
\affiliation{Department of Materials Engineering, Indian Institute of Science, 560012 Bangalore, India}

\begin{abstract}
Understanding the role of solute diffusivities in equilibrium tie-line selection during growth of a second phase in ternary 
and higher multicomponent two phase alloys is an important problem due to the strong dependence of mechanical 
properties on compositions. In this paper, we derive analytical expressions for predicting tie-lines 
and composition profiles in the matrix during growth of planar and cylindrical precipitates with 
the assumption of diagonal diffusivity matrices. 
We confirm our calculations by sharp interface and phase field simulations. The numerical  
techniques are in turn utilized for investigating the role of off-diagonal entries in the 
diffusivity matrix. In addition, the sharp interface methods allow for the tracking of 
the tie-line compositions during growth of 2D precipitates which contribute 
to an understanding of the change in equilibrium tie-lines chosen by the system during growth. 
\end{abstract}



\maketitle

\section{Introduction}
Properties of phases are closely related to their 
chemical compositions. In this respect, prediction of phase 
compositions during growth often becomes important with regards to 
the choice of alloy compositions and processing conditions. Theoretical understanding 
of the problem of diffusion controlled growth of phases from a super-saturated matrix in binary systems, 
is mainly due to the developments presented in~\cite{Zener1949,Frank1950,Ham1958}.  
While in binary alloys, the compositions of the phases are most of the time
determined directly from the phase diagram 
and the imposed temperature history, for ternary and higher 
multi-component alloys this becomes non-trivial as the 
two-phase equilibria is not unique. A consequence of this is
seen experimentally in Fe-C-Mo alloys~\cite{Bowman1946}, 
and other Fe-C-X systems~\cite{Bhadeshia1985,Aaronson1966} 
(where X stands for the substitutional alloying elements such as,  
Mn~\cite{Gilmour1972,Enomoto1987}, Ni~\cite{Sharma1973}, Cr~\cite{Sharma1979}) 
where the growth of a particular second-phase(ferrite) from the
matrix(austenite) occurs in the absence of any partition of the 
element X, which has very low diffusivity. This phenomenon,
which is also sometimes referred to as 'paraequilibrium', is 
a direct consequence of the existence of multiple equilibria
between the two-phases which is absent in binary alloys.
More pertinently, the choice of the equilibria can be shown to depend on the 
diffusivity of the different species as has already been 
shown in a classical theoretical analysis by Coates et al.~\cite{Coates1972,Coates1973},
where the conjugate problem of computing the bulk alloy compositions 
under different diffusivity matrices, for which 
a given tie-line is selected, is addressed. 
Another theoretical contribution by 
Bourne et al.~\cite{Bourne1994} derives expressions 
for composition profiles in the matrix along with predictions for 
equilibrium compositions. The method 
in~\cite{Bourne1994} is an indirect one, where the diffusivities are solved for 
by iterating over the possible tie-line compositions, whereby a 
prior knowledge of the equations of coexistence lines along with that of the 
tie-lines in the system is required.

This motivates our study in this paper which has three principal 
aims. Firstly, we  derive an analytical theory for diagonal 
diffusivity matrices (for both cylindrical and planar geometries), following previous
work in~\cite{Coates1972,Coates1973} and~\cite{Bourne1994}, that
not only allows a direct calculation of the quantities of interest 
(tie-lines and composition fields in the matrix) for a given bulk alloy composition 
and diffusivity matrix, but also incorporates the Gibbs-Thomson
correction of phase equilibria in 2D systems.
To this end, we have employed nothing more than the basic thermodynamic information 
associated with the variation of the free-energies of the phases 
with compositions, which differentiates it from the previous
approaches by presenting a more elegant way of prediction of
tie-lines given the bulk alloy composition and a diffusivity matrix which 
is easily extensible to an alloy system with any number of components.   

Secondly, a particular multi-component 
phase field model formulation based on a 
grand-potential formalism \cite{Choudhury+12, Plapp+11} 
is employed to study growth. The model itself has the possibility
to incorporate information from thermodynamic databases in an effective manner \cite{Choudhury+15}, 
which  will allow its subsequent utilization for
study of growth in real systems.
The phase field results are compared with both the 
analytical model and an independent sharp interface
(front tracking) numerical model, the objective being the validation of this
particular formulation for use in subsequent work involving
growth of multiple phases in complicated geometries.
In addition, the phase field and the sharp interface 
models are compared against each other for diffusivity
matrices comprising of off-diagonal entries, where 
an analytical treatment is difficult.
To our knowledge, although phase field models 
have been used for multi-component studies (Ti-Al-V~\cite{Chen2004}, 
Al-Si-Cu-Fe~\cite{Qin2005},  Al–Si–Cu–Mg–Ni~\cite{Bottger2006}, Mg-Al-Mn~\cite{Steinbach2007}, 
Ni-Al-Cr-Ta-W~\cite{Warnken2009}, Al-Si-Cu-Fe-Mg-Mn-Ni-Zn~\cite{Bottger2010}, Fe-C-B~\cite{Lentz2015}), 
the influence of diffusivity matrices on the choice of equilibria
has not been dealt in detail. For example in \cite{Yeon2001},
the authors do investigate the particular case of `paraequilibrium' 
in the case of an Fe-C-Mn alloy, however, the simulations are 
not directed towards deriving the phase compositions in the 
relevant scaling regime during growth. 

Thirdly, as a result of the comparison of the different methods
and the investigation of the dependence of the phase equilibria on the 
diffusivity/mobility matrices we highlight 
the need for their accurate measurement, 
without which the implications of understanding
derived from numerical simulations, such as the phase field
method, become unreliable. 

We begin our discussion with planar growth in Section II 
where a theoretical analysis for the case of diagonal diffusivities 
in the scaling regime is presented in Section IIA, followed
by a description of the numerical simulation methods.
The sharp interface (front tracking model) is 
described in Section IIB and phase field (diffuse interface mode) 
in Section IIC (that is generic for all dimensions), 
which utilize the thermodynamics of a 
representative alloy system elaborated in Section IID.
Subsequently, we present the comparison between the 
analysis and the simulations for planar growth in Section IIE.
Thereafter, radial growth is analyzed in Section III, 
with the theoretical development in Section IIIA and
the description of the corresponding numerical sharp-interface model 
for radial growth in Section IIIB, followed by the comparison
of the composition profiles in the matrix between analysis
and simulation methods in Section IIIC. We end with a discussion 
in Section IV and conclusions and outlook in Section V.

\section{Planar growth}
In this section, we develop an analytical theory and conduct sharp interface and phase field simulations to 
understand the problem of tie-line selection during planar growth of a multicomponent alloy.  

\subsection{Theory}
In the theoretical analysis, we are going to consider
the situation where we have a 1D domain, with one-sided
diffusivity, i.e. only the matrix has non-zero diffusion 
coefficients. Additionally, we will restrict our theoretical
discussion to diagonal diffusivities. Furthermore, 
to ensure brevity, we express all the equations in the vector-matrix notation, 
where $\left\lbrace\cdot\right\rbrace$ represents a vector and 
$\left[\cdot\right]$ represents a matrix.
We start with the following governing equations in the matrix
which write as, 

\begin{align}
 \left\lbrace\dfrac{\partial c_i}{\partial t}\right\rbrace &= 
 \left[D_{ij}\right]\left\lbrace\dfrac{\partial^{2}c_j}{\partial x^{2}}\right\rbrace, 
 \label{non-normalized-governing-equations}
\end{align}

where, both the indices $i$ and $j$ iterate over all the 
($K-1$) solute components in a $K$ component
system. The Stefan boundary condition at the interface writes as, 

\begin{align}
 v\left\lbrace c_{i,eq}^{\alpha} - c_{i,eq}^{\beta}\right\rbrace &= 
 -\left[D_{ij}\right]\left\lbrace\dfrac{\partial c_j}{\partial x}\right\rbrace\Bigg|_{x_f},
 \label{non-normalized-stefan-condition}
\end{align}

with $c_{i,eq}^{\alpha, \beta}$ as the equilibrium compositions 
at the interface for the matrix ($\alpha$) and the precipitate ($\beta$) and $v$
being the velocity of the interface with its position being denoted by $x_f$. 

Here, we restrict ourselves to the situation corresponding to independent solute diffusion 
and perform a co-ordinate transformation, writing 
$\eta = x/\sqrt{t}$. In these co-ordinates the 
governing equations transform to, 

\begin{align}
 \left\lbrace\dfrac{\partial^{2}c_i}{\partial \eta^{2}}\right\rbrace 
 &= \dfrac{-\eta}{2}\left\lbrace\dfrac{1}{D_{ii}}\dfrac{\partial c_i}{\partial \eta} \right\rbrace,
 \label{governing_equations}
\end{align}

while the Stefan-boundary condition at the interface 
reads, 

\begin{align}
 \left\lbrace\dfrac{\partial c_i}{\partial \eta}\right\rbrace\Bigg|_{\eta_s} 
 &= \dfrac{-\eta_s }{2}\left\lbrace\dfrac{\Delta c_i}{D_{ii}}\right\rbrace,
 \label{Stefan_condition}
\end{align}

where $\eta_s = x_f/\sqrt{t}$ is the corresponding value at the interface, 
which is at a position $x_f$ at a given time $t$. We have used,
$\Delta c_i=\left(c_{i,eq}^{\alpha} - c_{i,eq}^{\beta}\right)$.

Integrating the Eq. \ref{governing_equations} once, we derive,

\begin{align}
 \left\lbrace\dfrac{\partial c_i}{\partial \eta}\right\rbrace 
 &= \left\lbrace\lambda_i \exp \left(\dfrac{-\eta^{2}}{4D_{ii}}\right) \right\rbrace,
 \label{integrate_once}
\end{align}

where $\lambda_i$'s are integration constants. Using the Stefan's conditions in 
Eq.~\ref{Stefan_condition}, the value of the integration constants can 
be derived as, 

\begin{align}
 \left\lbrace\lambda_i\right\rbrace &= 
 \dfrac{-\eta_s}{2}\left\lbrace\dfrac{\Delta c_i}{ D_{ii} \exp \left(\dfrac{-\eta_s^{2}}{4 D_{ii}}\right)}\right\rbrace,
\end{align}

where, since $\lambda_i$'s are independent of $\eta$, the value of $\eta_s$ must 
be a constant for the given alloy composition.

Integrating the Eq.~\ref{integrate_once} once again and invoking
the boundary condition that the far-field compositions corresponding
to $\eta=\infty$ are known as $\left\lbrace c_i^{\infty}\right\rbrace$ and the compositions
at the interface $\eta=\eta_s$ are the equilibrium compositions 
$\left\lbrace c_{i,eq}^{\alpha}\right\rbrace$, we derive, 

\begin{align}
 \left\lbrace\dfrac{c_i^{\infty} - c_{i,eq}^{\alpha}}{\int_{\eta_s}^{\infty}\exp\left(-\dfrac{\eta^{2}}{4D_{ii}}\right)d \eta} \right\rbrace
 &= \dfrac{-\eta_s}{2}\left\lbrace \dfrac{\Delta c_i}{D_{ii} \exp \left(\dfrac{-\eta_s^{2}}{4D_{ii}}\right)}\right\rbrace,
\end{align}

which can be reduced using the complement of the error functions as, 

\begin{align}
 \left\lbrace c_i^{\infty} - c_{i,eq}^{\alpha} \right \rbrace &= 
 -\dfrac{\sqrt \pi \eta_s}{2}\left\lbrace \dfrac{\Delta c_i \erfc \left(\dfrac{\eta_s}{2\sqrt{D_{ii}}}\right)}{\sqrt{D_{ii}} \exp \left(\dfrac{-\eta_s^{2}}{4D_{ii}}\right)}\right\rbrace,
 \label{composition_profiles}
\end{align}

\begin{figure}[!htbp]
 \includegraphics[width=\linewidth]{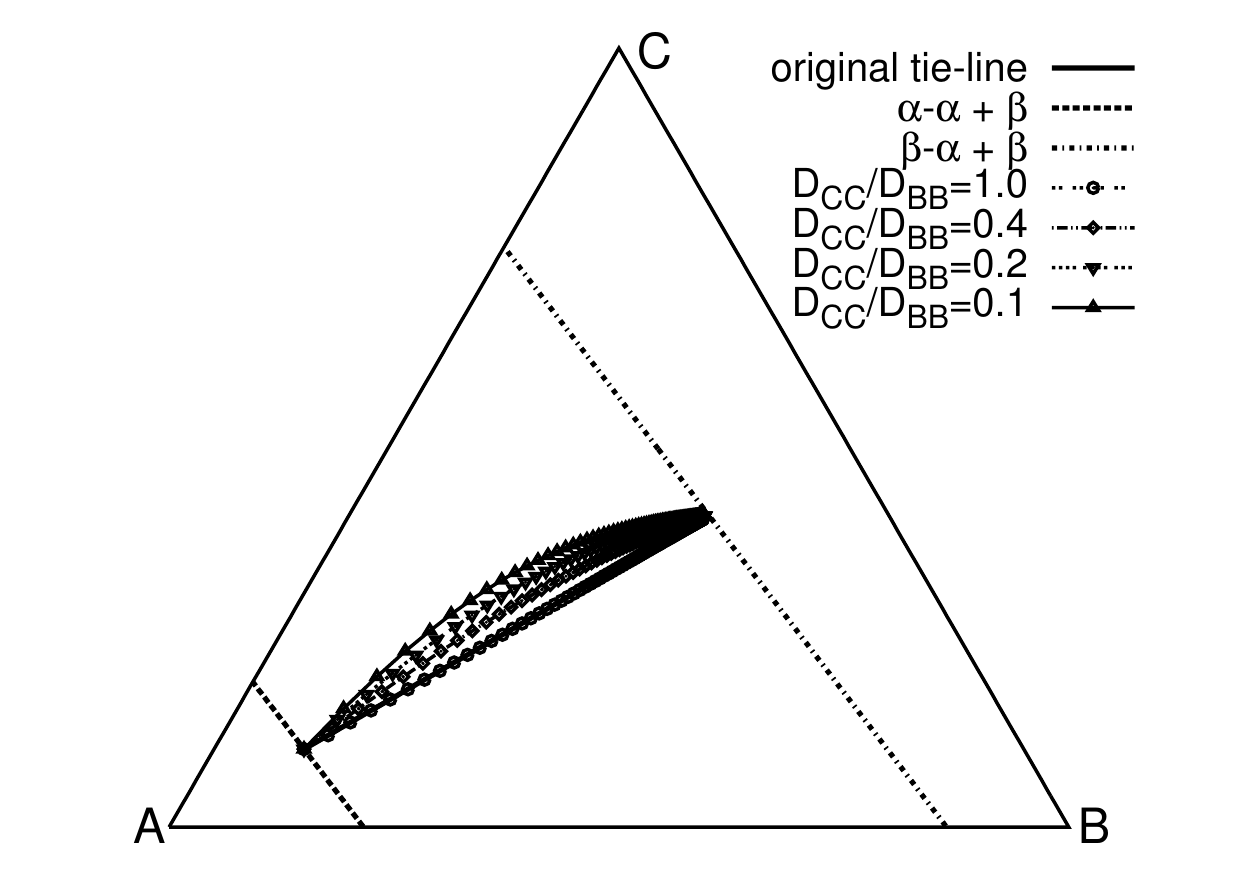}
 \caption{Loci of alloy compositions which give the same tie-line given by the equilibrium
 matrix ($\alpha$) compositions ($c_B,c_C$) as ($0.1,0.1$) and precipitate ($\beta$) composition ($0.4,0.4$)
 for different diffusivity ratios $D_{CC}/D_{BB}$.}
 \label{Alloy_compositions_diff_DAA_DBB}
\end{figure}

In the following, we apply the above analysis for a three component system
where $B$, $C$ are the solutes, and the values $c_{B,eq}^{\alpha}$,
$c_{C,eq}^{\alpha}$ refer to the equilibrium compositions of the two components
in the matrix phase at the interface. Eq.~\ref{composition_profiles} 
can be utilized in two ways, the first is to derive all the alloy 
compositions $c_B^{\infty}$ and $c_C^{\infty}$, that can have a given tie-line one of whose 
ends is given by the matrix composition
at the interface, which is highlighted
in Fig.~\ref{Alloy_compositions_diff_DAA_DBB}. One can see that for 
the situation where $D_{BB}/D_{CC}$ is unity, the corresponding
tie-lines for all values of $\eta_s$, are equal to the thermodynamic tie-line
containing the alloy composition. With change in the ratio of $D_{BB}/D_{CC}$,
the alloy compositions start to shift from the thermodynamic tie-line
and the graph portrays all the possibilities which are derived through 
variation of the value of $\eta_s$, each for a different ratio of 
$D_{BB}/D_{CC}$.

\begin{figure}[!htbp]
 \includegraphics[width=0.4\textwidth]{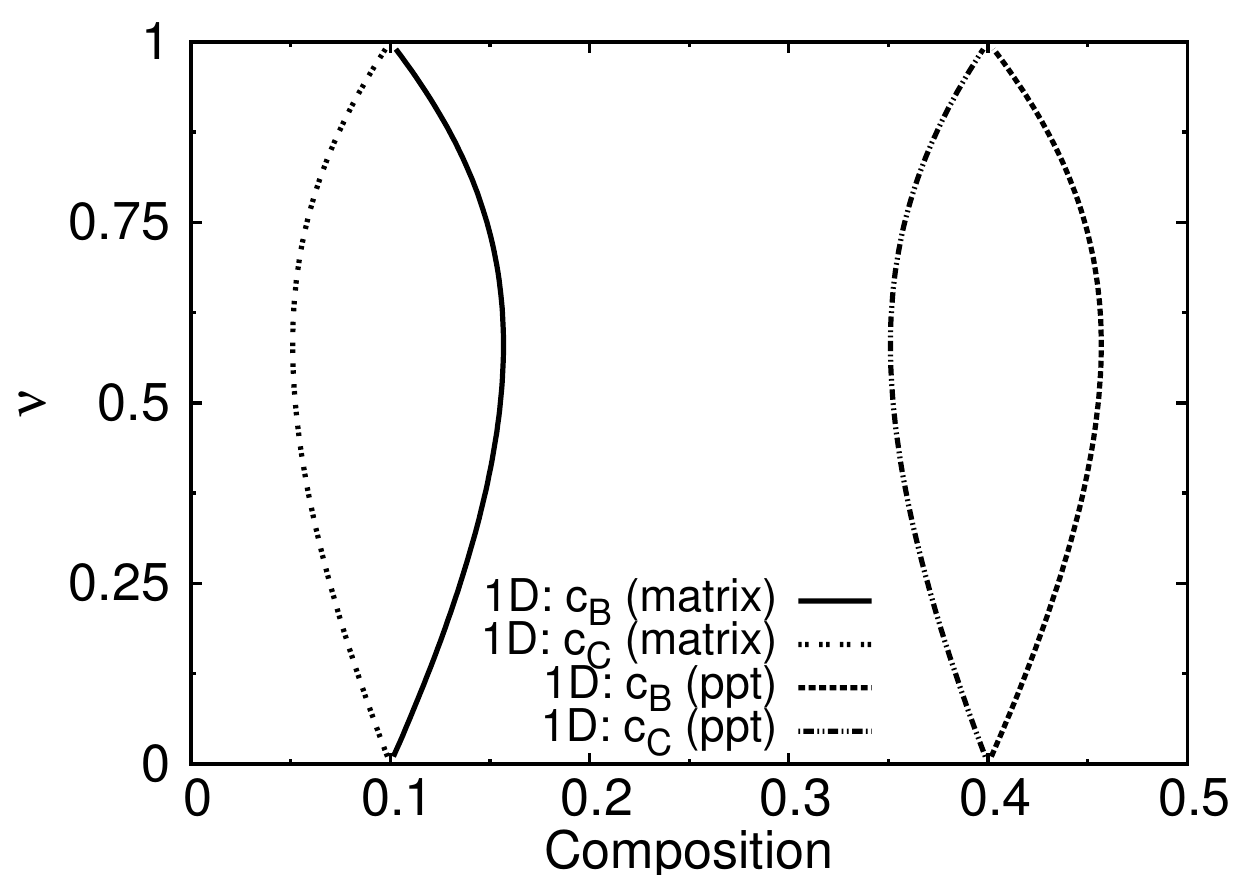}
 \caption{Combination of tie-line equilibrium compositions ($c_B,c_C$) for the matrix
 and the precipitate corresponding to different alloy compositions along the tie-line ($0.1,0.1$) (matrix)
 ($0.4,0.4$) (precipitate) compositions, for a ratio of diffusivity $D_{CC}/D_{BB}=0.1$.
 The thermodynamic information about co-existence lines is derived from 
 Eq.~\ref{chemical_potential_coexistence}.
 The parameter $\nu$ signifies a point corresponding to the 
 choice along this tie-line; $\nu=0$ represents an alloy composition 
 equal to the matrix ($\alpha$) while $\nu=1$ for an alloy composition equal 
 to that of the precipitate ($\beta$).}
 \label{Tie-lines}
\end{figure}

The second possibility is to derive the equilibrium compositions $c_{B,eq}^{\alpha}$ 
and $c_{C,eq}^{\alpha}$, and the value of $\eta_s$ using Eq.~\ref{composition_profiles}, 
given the bulk alloy composition $c_B^{\infty}$ and $c_C^{\infty}$ and the diffusivity matrix.
This however requires that we know the functional relationships $c_B\left(\mu_B,\mu_C\right)$
and $c_C\left(\mu_B,\mu_C\right)$ and also the relationship between
$\mu_{B,eq}\left(\mu_{C,eq}\right)$ which is a property of the 
thermodynamic co-existence line. Using them, the two equations in 
(\ref{composition_profiles}) can be reduced to a system of two 
equations containing $\mu_{C,eq}$ and $\eta_s$, which can 
then be consistently solved for. The resulting $\mu_{C,eq}$
can then be utilized to fix $\mu_{B,eq}$ using the equilibrium 
thermodynamics of the co-existence line and thereby the matrix compositions 
$c_B^{\alpha}\left(\mu_B,\mu_C\right)$, $c_C^{\alpha}\left(\mu_B,\mu_C\right)$.
Further, the precipitate compositions can also be fixed using
the corresponding relations for $c_B^{\beta}\left(\mu_B,\mu_C\right)$, 
$c_C^{\beta}\left(\mu_B,\mu_C\right)$.
For a linearized phase diagram we delineate the 
different possible tie-lines for each of 
the bulk alloy compositions for a single diffusivity ratio 
$D_{BB}/D_{CC}$ in Fig.~\ref{Tie-lines}.

The composition profile for a given $\eta_s$ and far-field composition 
can be derived simply by using the Eq.~\ref{composition_profiles}
as,

\begin{align}
 \left\lbrace c_i\left(\eta\right)\right\rbrace 
 = \left\lbrace c_i^{\infty} - \left(c_i^{\infty} 
 - c_{i,eq}^{\alpha}\right)\dfrac{\erfc\left(\dfrac{\eta}{2\sqrt{D_{ii}}}\right)}{\erfc\left(\dfrac{\eta_s}{2\sqrt{D_{ii}}}\right)}\right\rbrace,
\label{1D_comp_profile}
 \end{align}

where $\eta = x/\sqrt{t}$ is the variable
in the transformed co-ordinate system, for all $x>x_f$
and time $t$, $x_f$ being the position of the interface.

In the following two subsections, we describe two numerical 
simulation models namely, a front-tracking (sharp interface) model
and a diffuse interface (phase field) model, 
for treating the transient evolution of the interface
under a general condition of arbitrary diffusivity matrices.

\subsection{Sharp interface model}
The governing Eq.~\ref{non-normalized-governing-equations}
can be solved numerically along with the boundary condition at the 
interface given by Eq.~\ref{non-normalized-stefan-condition} using an 
advective scheme where the motion of the interface is accounted for 
by an equal and opposite advective term which brings the interface 
back to its origin place (this is a numerically effective way
of treating the front tracking problem as the interface cell 
remains invariant). Thereby, the interface stays stationary in 
this frame of reference. The modified governing equations in the vector-matrix notation are written 
as, 



\begin{equation}
 \left\lbrace\dfrac{\partial c_i}{\partial t} -v\dfrac{\partial c_i}{\partial x}\right\rbrace = 
 \left[D_{ij}\right]\left\lbrace\dfrac{\partial^2 c_j}{\partial x^2}\right\rbrace,
 \label{advective-governing-equation}
\end{equation}

where $v$ is the instantaneous local velocity of the interface.
The position of the interface is marked by the interface 
cell, where the equilibrium compositions
of the precipitate and the matrix are specified. 
For a ternary system with $B$ and $C$ as solutes, 
the gradients $\partial c_{B,C}/\partial x$ are written 
in discrete form using these compositions at the interface 
cell and the bulk compositions on the matrix side, next to the interface.
To compute the interfacial compositions, 
the pair of equations in~(\ref{non-normalized-stefan-condition}) 
are self-consistently solved for both the velocity and the 
equilibrium chemical potentials $\mu_{B,eq}, \mu_{C,eq}$ at
the interface. This requires that the composition functions 
$c_{B,C}^{\alpha,\beta}\left(\mu_B, \mu_C\right)$ are known
from thermodynamics along with the equilibrium relation 
between the two chemical potentials,
$\mu_{B,eq}\left(\mu_{C,eq}\right)$. 

Using these, the pair of equations in~(\ref{non-normalized-stefan-condition}),
can be transformed
such that the unknowns are the equilibrium
chemical potential of a given component ($\mu_{C,eq}$) and the velocity ($v$).
Once the chemical potential $\mu_{C,eq}$ is known, the other 
chemical potential $\mu_{B,eq}$ can also be fixed using the 
thermodynamic relation between them and thereby also 
the individual phase compositions at the interface,
$c_{B,C}^{\alpha,\beta}\left(\mu_B, \mu_C\right)$. 
The compositions are then utilized to compute the gradients 
in the governing equations in Eq. \ref{advective-governing-equation}
and thereby evolve the compositions $c_{B,C}$ in the matrix
phase through one time-step, using also the velocity 
that is derived from the Stefan condition. When the scaling
regime is reached, the equilibrium compositions at
the interface reach a steady state. The corresponding
composition profiles can then be compared with the 
theoretical analysis in the preceding section. 
This method, gives a fast accurate
benchmark in 1D, in comparison to the more computationally
intensive method to be discussed in the following 
section.

\subsection{Phase field model}
In this subsection, we give a brief description of 
phase field (diffuse interface) model, also described in \cite{Choudhury+11-3, Choudhury+12},
for studying growth in multi-component systems.
All the equations are presented in
the tensorial form and so they describe the model regardless of the dimensionality considered.
Phase evolution is determined by the phenomenological
minimization of the grand
canonical density functional ($\Omega$) written as,
\begin{align}
 {\Omega}\left(\vmu,T,\phi\right)&=\int_{V}\Bigg[\Psi\left(\vmu,T,
\phi\right) + \nonumber\\ 
&\left(\epsilon a\left(\phi,\nabla \phi\right) +
\dfrac{1}{\epsilon}w\left(\phi\right)\right)\Bigg]dV,
 \label{GrandPotentialfunctional}
\end{align}
where $\epsilon$ is the length scale related to the diffuse interface,
and $\phi$ is the order parameter determining the presence of
the precipitate phase, i.e., regions where $\phi=1$, demarcate the precipitate,
and $\phi=0$, the matrix. The functional $a\left(\phi,\nabla \phi\right)$ is the gradient 
energy written as $\sigma|\nabla \phi|^{2}$, $\sigma$ being numerically
equal to the interfacial energy. Functional $w\left(\phi\right)$ is a surface 
potential density written as a double-well function of the order-parameter $\phi$,
which is $9\sigma \phi^{2}\left(1-\phi\right)^{2}$. We write the grand potential density 
$\Psi$ as an interpolation of the individual grand 
potential densities $\Psi^\alpha, \Psi^\beta$, each of which 
are functions of the diffusion potential vector 
$\vmu=\lbrace{\mu_1,\ldots,\mu_{K-1}\rbrace}$ 
containing the $K-1$ independent diffusion potentials and 
temperature T in the system as,
\begin{align}
 \Psi\left(\vmu,T,\phi\right) &= \Psi^{\alpha}\left(\vmu,T\right)h\left(1-\phi\right) \nonumber\\ 
                               &+ \Psi^{\beta}\left(\vmu,T\right)h\left(\phi\right)
\label{GP_interpolation}
\end{align}

$h\left(\phi\right)$ is the interpolation polynomial
written as $h\left(\phi\right) = \phi^{2}\left(3-2\phi\right)$, 
which ensures that $h\left(\phi\right) + h\left(1-\phi\right)=1$.
The phase concentration expressions 
can be derived in terms of the diffusion potential, using the relation,
\begin{align}
c_{i}^{\alpha,\beta}\left(\vmu, T\right)=
-V_m\dfrac{\partial \Psi^{\alpha,\beta}\left(\vmu,T\right)}{\partial \mu_i},
\label{cofmu}
\end{align}
where $V_m$ is the molar volume which is considered equal for all
the components in this entire paper.
Thereafter, the equations of motion for the phase field 
and the composition variables are derived in a standard manner.

The evolution equations for the phase field $\phi$ can be derived as,
\begin{align}
\tau_{\alpha\beta} \epsilon \dfrac{\partial \phi}{\partial t}=\epsilon \left(\nabla
\cdot \dfrac{\partial a\left(\phi,\nabla \phi\right)}{\partial \nabla
\phi}\right)-\dfrac{1}{\epsilon}\dfrac{\partial
w\left(\phi\right)}{\partial \phi}-\dfrac{\partial
\Psi\left(\vmu, T, \phi\right)}{\partial \phi},
\label{Equation6_grandchem}
\end{align}
where, the relaxation constants for the matrix-precipitate interfaces 
$\tau_{\alpha \beta}$, are calculated for
pure diffusion-controlled regime using the analysis described
in \cite{Choudhury+12}.

For a general multi-phase, multi-component system, 
the evolution equations for the components of the diffusion potential
$\vmu$ can be expressed in vector-matrix form by,

\begin{align}
 &\left\lbrace\dfrac{\partial \mu_i}{\partial t}\right\rbrace = \nonumber\\
&\left[\sum^{p=\alpha,\beta} 
h_p\left(\phi\right)\dfrac{\partial c_i^p\left(\vmu,
T\right)}{\partial \mu_j}\right]^{-1}_{ij}\Big\lbrace\nabla\cdot\sum_{j=1}^{K}M_{ij}
\left(\phi\right)\nabla \mu_j \nonumber\\ 
&- \sum^{p=\alpha,\beta} c^p_{i}\left(\vmu,T\right)\dfrac{\partial
h_p\left(\phi\right)}{\partial t}\Big\rbrace.
\label{Mu_explicit_temperature}
\end{align}

where $\left\lbrace \cdot \right\rbrace$ represents a vector 
of dimension $(K-1)$ while $\left[\cdot\right]$ denotes 
a matrix of dimension $(K-1) \times (K-1)$.
For conciseness, we have utilized expressions $h_\alpha\left(\phi\right)=h\left(1-\phi\right)$
and $h_\beta\left(\phi\right)= h\left(\phi\right)$.
Here, $M_{ij}\left(\phi\right)$ is the atomic mobility, where the
individual phase mobilities are interpolated as,
\begin{align}
 M_{ij}\left(\phi\right) &= M_{ij}^\alpha(1-\phi) + M_{ij}^\beta\phi.
 \label{mobility_interp}
\end{align}
Each of the $M_{ij}^{\alpha,\beta}$ is defined using the expression,
\begin{align}
 \left[M_{ij}^{\alpha,\beta}\right] &= \left[D^{\alpha,\beta}_{ik}\right] \left[\dfrac{\partial
c_k^{\alpha,\beta}\left(\vmu,T\right)}{\partial \mu_j}\right],
\label{mobility}
\end{align}
where $D_{ij}^\alpha$ and $D_{ij}^\beta$ are the solute inter-diffusivities in $\alpha$ and $\beta$ respectively.

For the simulations in the present section, we will impose
diffusivities only in one of the phases (matrix phase, $\alpha$), which
implies the diffusivity matrix is zero for the precipitate
phase, $\beta$. The anomalous artificial solute trapping that is known 
to arise because of this choice is countered by using a multi-component
version of the anti-trapping current that is derived in 
\cite{Choudhury+12}. This is an additional flux in the diffusion 
equation which acts towards the matrix phase.

\subsection{Thermodynamics: Equilibrium across a planar front}
The driving forces for a phase transformation from 
$\alpha$ to $\beta$ is the difference between the grand-potential densities
of the phases, i.e., $\Delta\Psi=\Psi^{\alpha}-\Psi^{\beta}$. 
In this section we utilize a linearized phase diagram around
the compositions of interest, which also allows for a simple 
coupling to thermodynamic databases.
The driving force, $\Delta \Psi$ is derived by linearly expanding the 
individual grand-potential densities in terms of
the departure of the diffusion potential from a given equilibrium 
value $\vmu_{eq}^{*}=\left\lbrace\mu_{i,eq}^{*}\right\rbrace$ as,

\begin{align}
 \Psi^{\alpha,\beta}\left(\vmu, T\right) &= \Psi^{\alpha,\beta}\left(\vmu_{eq}^{*}, T\right)+ \nonumber\\
 &\left\lbrace\dfrac{\partial\Psi^{\alpha,\beta}}{\partial\mu_i}\right\rbrace_{\mu_{i,eq}^{*}}\left\lbrace\mu_i-\mu_{i,eq}^{*}\right\rbrace
 \label{driving_force}
\end{align}

and therefore, the leading order term in the driving force for the phase transformation 
$\alpha$ to $\beta$ writes as, 

\begin{align}
 \Delta\Psi^{\alpha\beta} &= \left(\Psi^{\alpha}-\Psi^{\beta}\right)\nonumber\\
                          &=  \left\lbrace\dfrac{\partial\Psi^{\alpha}}{\partial\mu_i} - 
                          \dfrac{\partial\Psi^{\beta}}{\partial\mu_i}\right\rbrace_{\mu_{i,eq}^{*}}\left\lbrace\mu_i-\mu_{i,eq}^{*}\right\rbrace,
\end{align}

where, we have used the vector-matrix notation. 
Using the thermodynamic relationship in in Eq.~\ref{cofmu}, 
we can equivalently write the preceding equation as, 

\begin{align}
 \Delta\Psi^{\alpha\beta}=  \dfrac{1}{V_m}\left\lbrace c_{i,eq}^{\beta,*} - c_{i,eq}^{\alpha,*}\right\rbrace \left\lbrace\mu_i-\mu_{i,eq}^{*}\right\rbrace.
 \label{linearized_driving_force}
\end{align}


The phase compositions as a function of the chemical potential 
are thereafter linearly extrapolated from the chosen 
equilibrium points $\left\lbrace c_{i,eq}^{*}\right\rbrace$ as, 
\begin{align}
  \left\lbrace c_{i}^{\alpha,\beta}\right\rbrace &= \left\lbrace c_{i,eq}^{\alpha,\beta}\right\rbrace^{*} 
  + \left[\dfrac{\partial c_i^{\alpha,\beta}}{\partial \mu_j}\right]_{\mu_{i,eq}^{*}}\left\lbrace \mu_j - \mu_{j,eq}^{*}\right\rbrace,
  \label{linear_composition_profile}
\end{align}

which are utilized in the evolution equation of the components as described in 
Eq.~\ref{Mu_explicit_temperature}. The susceptibility matrix
$\left[\dfrac{\partial c_i^{\alpha,\beta}}{\partial \mu_j}\right]_{\mu_{i,eq}^{*}}$ which is
also used in the construction of the atomic mobility matrix in Eq.~\ref{mobility} is 
a term that can be also retrieved from the thermodynamic databases. 
The equilibrium composition vectors $\left\lbrace c^{\alpha,\beta}_{eq}\right\rbrace^{*}$ 
of the phases are used in constructing a linearized expansion of the driving force, as
given in Eq.~\ref{linearized_driving_force} and eventually in the evolution equation 
of the order-parameter given in Eq.~\ref{Equation6_grandchem} (all quantities which are denoted 
with a superscript $^{*}$, pertain to values around which the linearization is performed).
\footnote{Strictly speaking, the composition relation in Eq. \ref{linear_composition_profile} is 
derived from a grand-potential density which is parabolic in $\vmu$. However, in 
this particular paper, we utilize a linearized form of the driving force in order
to simplify its utilization in several non-linear equations in the sharp interface 
and analytical models. This just modifies the thermodynamic relation between
the equilibrium diffusion potentials.} 

\begin{figure}[!htbp]
\begin{center}
\begin{tikzpicture}[domain=2:4]

\draw [color=red,thick,-] ($(0,0)+(1.0,0)$) 
    arc (0:60:1.0);
\draw [color=blue,thick,-] ($(0,0)+(2.0,0)$) 
    arc (0:60:2.0);
        
\pgfmathsetmacro{\XValue}{2.0*cos(20)}%
\pgfmathsetmacro{\YValue}{2.0*sin(20)}%

\pgfmathsetmacro{\xValue}{cos(20)}%
\pgfmathsetmacro{\yValue}{sin(20)}%

\draw [thin, solid] ($(\xValue,\yValue)$) -- ($(\XValue,\YValue)$);

\pgfmathsetmacro{\XValue}{2.0*cos(10)}%
\pgfmathsetmacro{\YValue}{2.0*sin(10)}%

\pgfmathsetmacro{\xValue}{cos(10)}%
\pgfmathsetmacro{\yValue}{sin(10)}%

\draw [thin, solid] ($(\xValue,\yValue)$) -- ($(\XValue,\YValue)$);

\pgfmathsetmacro{\XValue}{2.0*cos(30)}%
\pgfmathsetmacro{\YValue}{2.0*sin(30)}%

\pgfmathsetmacro{\xValue}{cos(30)}%
\pgfmathsetmacro{\yValue}{sin(30)}%

\draw [thin, solid] ($(\xValue,\yValue)$) -- ($(\XValue,\YValue)$);

\pgfmathsetmacro{\XValue}{2.0*cos(40)}%
\pgfmathsetmacro{\YValue}{2.0*sin(40)}%

\pgfmathsetmacro{\xValue}{cos(40)}%
\pgfmathsetmacro{\yValue}{sin(40)}%

\draw [thin, solid] ($(\xValue,\yValue)$) -- ($(\XValue,\YValue)$);

\pgfmathsetmacro{\XValue}{2.0*cos(50)}%
\pgfmathsetmacro{\YValue}{2.0*sin(50)}%

\pgfmathsetmacro{\xValue}{cos(50)}%
\pgfmathsetmacro{\yValue}{sin(50)}%

\draw [thin, solid] ($(\xValue,\yValue)$) -- ($(\XValue,\YValue)$);

\pgfmathsetmacro{\xValue}{cos(30)}%
\pgfmathsetmacro{\yValue}{sin(30)}%

\pgfmathsetmacro{\XValue}{2.0*cos(30)}%
\pgfmathsetmacro{\YValue}{2.0*sin(30)}%

\fill[red] ($(\xValue,\yValue)$) circle(2.5pt);
\fill[blue] ($(\XValue,\YValue)$) circle(2.5pt);

\pgfmathsetmacro{\XValue}{1.38*cos(30)*cos(60)}%
\pgfmathsetmacro{\YValue}{1.38*cos(30)*sin(60)}%

\pgfmathsetmacro{\xValue}{1.38*cos(30)}%
\pgfmathsetmacro{\yValue}{0}%

\draw [thin, solid] ($(\xValue,\yValue)$) -- ($(\XValue,\YValue)$);

\pgfmathsetmacro{\XValue}{2.7*cos(30)*cos(60)}%
\pgfmathsetmacro{\YValue}{2.7*cos(30)*sin(60)}%

\pgfmathsetmacro{\xValue}{2.7*cos(30)}%
\pgfmathsetmacro{\yValue}{0}%
\draw [thin, solid] ($(\xValue,\yValue)$) -- ($(\XValue,\YValue)$);

\pgfmathsetmacro{\xValue}{1.5*cos(30)}%
\pgfmathsetmacro{\yValue}{1.5*sin(30)}%

\draw[mark=square*,mark size=2.5pt,mark options={color=black}]  plot coordinates {($(\xValue,\yValue)$)};

\pgfmathsetmacro{\xValue}{0.5*cos(30)}%
\pgfmathsetmacro{\yValue}{0.5*sin(30)}%

\pgfmathsetmacro{\XValue}{2.75*cos(30)}%
\pgfmathsetmacro{\YValue}{2.75*sin(30)}%

\node[color=red] at  ($(\xValue, \yValue)$){$\alpha$};
\node[color=blue] at ($(\XValue, \YValue)$){$\beta$};

\draw[black] (0,0)--(4,0);
\draw[black] (0,0)--(0.5*4,1.732/2 * 4);
\draw[black] (0.5*4,1.732/2 *4)--(4,0);

\end{tikzpicture}
\end{center}
\caption{Two-phase equilibrium in a ternary alloy. The phase-coexistence lines at a given temperature 
are drawn as solid curves and the corresponding tie-lines are drawn as solid black lines between the
co-nodes on the co-existence curves. For a particular alloy composition indicated by the solid square, the 
susceptibility matrix is computed corresponding to the compositions of the phases comprising the 
tie-line, which are marked here by solid circles and which are also the 
composition vectors $\vc^{\alpha,*}_{eq},\vc^{\beta,*}_{eq}$ that are used in the approximation. 
The tangents to the co-existence tie-lines are the local extrapolations corresponding
to the these thermodynamic properties of the alloy at the respective phase compositions.}
\label{Tie_line_assumption}
\end{figure}
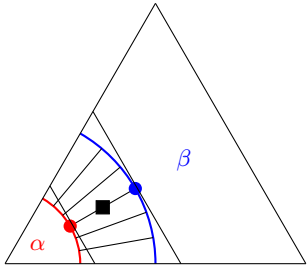

Using the approximate driving forces as the leading order term in the expansion of the grand-potential
densities, also fixes the relation between the diffusion potentials of the different 
components along the co-existence lines, which is derived by 
setting, $\Delta\Psi^{\alpha\beta}=0$ in Eq.\ref{linearized_driving_force}. 
For the case of a ternary alloy, this relation reads,

\begin{align}
 \dfrac{\mu_{B,eq}-\mu_{B,eq}^{*}}{c_{C,eq}^{\alpha,*} - c_{C,eq}^{\beta,*}} 
 = -\dfrac{\mu_{C,eq}-\mu_{C,eq}^{*}}{c_{B,eq}^{\alpha,*} - c_{B,eq}^{\beta,*}}
 \label{chemical_potential_coexistence}
\end{align}

i.e, the set of equilibrium diffusion potentials of the components
are related to each other using the previous relation. The equilibrium
phase co-existence lines can also be derived after some algebraic
manipulation as,

\begin{align}
 \dfrac{c_{B,eq}^{\alpha,\beta} - \left(c_{B,eq}^{\alpha,\beta}\right)^{*}}{c_{C,eq}^{\alpha,\beta} - \left(c_{C,eq}^{\alpha,\beta}\right)^{*}} &= 
 \dfrac{\left(\dfrac{\partial c_B^{\alpha,\beta}}{\partial \mu_C} - \dfrac{1}{\rho}\dfrac{\partial c_B^{\alpha,\beta}}{\partial \mu_B}\right)^{*}}
 {\left(\dfrac{\partial c_C^{\alpha,\beta}}{\partial \mu_C} - \dfrac{1}{\rho}\dfrac{\partial c_C^{\alpha,\beta}}{\partial \mu_B}\right)^{*}},
 \label{approximate_coexistence}
\end{align}

where $\rho =(c_{B,eq}^{\alpha,*} - c_{B,eq}^{\beta,*})/(c_{C,eq}^{\alpha,*} - c_{C,eq}^{\beta,*})$.
Therefore, given a set of equilibrium compositions $c_{B,eq}^{\alpha,*},c_{C,eq}^{\alpha,*},c_{B,eq}^{\beta,*},c_{C,eq}^{\beta,*}$,
and the corresponding susceptibility $\left[\dfrac{\partial c_i^{\alpha,\beta}}{\partial \mu_j}\right]_{\mu_{j,eq}^{*}}$, 
which are two quantities that can be derived from thermodynamic
databases, the equilibrium co-existence lines in the vicinity of the chosen compositions are
correctly represented for the given system of interest. 
The susceptibility matrix can be easily determined
by computing the inverse of the matrix, 
$\left[\dfrac{\partial \mu_{i}^{\alpha,\beta}}{\partial c_j}\right]_{\left(c_{j,eq}^{\alpha,\beta}\right)^{*}}$,
which can be retrieved from the derivatives of the free-energy expressions in the databases
near the compositions of interest. Fig.~\ref{Tie_line_assumption} sketches the approximate
scheme that is used and the resulting co-existence lines corresponding to the 
expressions in Eq.~\ref{approximate_coexistence}.

\subsection{Results}
Having described the analytical theory and the numerical techniques 
that will be employed to understand precipitate growth in a ternary alloy in 1D,  
we begin this section by comparing predictions of composition profiles from 
phase field, sharp interface and analytical expression against each other\footnote{An important point to note 
here is that our analytical expressions are derived for a system where the precipitate phase 
grows into a supersaturated matrix which is infinite in extent. To ensure that our numerical calculations,
in spite of being performed on finite systems, are representative of infinite systems,
necessitate a few modifications. 
Phase field calculations are performed in a simulation box attached to the diffuse interface,
which allows us to capture the scaling regime, otherwise impeded 
by the change in far-field compositions at the system boundaries due to growth,
while sharp interface calculations are performed by expressing the governing equations in a reference frame
attached to the interface. Thus, both our numerical techniques are attuned to capture growth in an infinite system.}. 
In this regard, it's important to note that all our calculations are performed in the non-dimensional setting. 
The definitions of the relevant scales that can be used to convert dimensionless quantities
into dimensional values for a particular system are presented in the Appendix.

\subsubsection{Three-way comparison between analytical theory and simulation methods}

\begin{figure}[!htbp]
\centering
\subfigure[]{\includegraphics[width=0.4\textwidth]{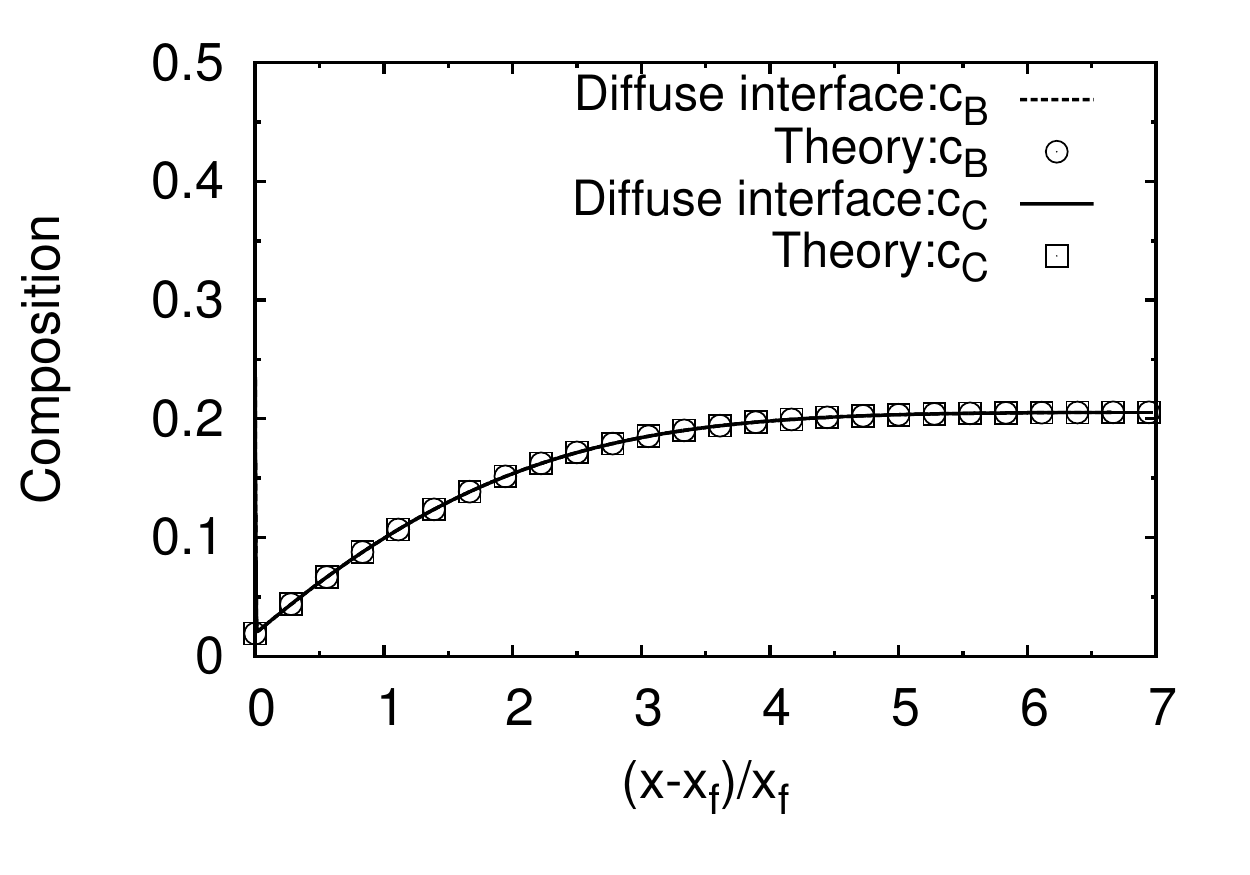}
\label{identD_t_l_known}
}
\subfigure[]{\includegraphics[width=0.4\textwidth]{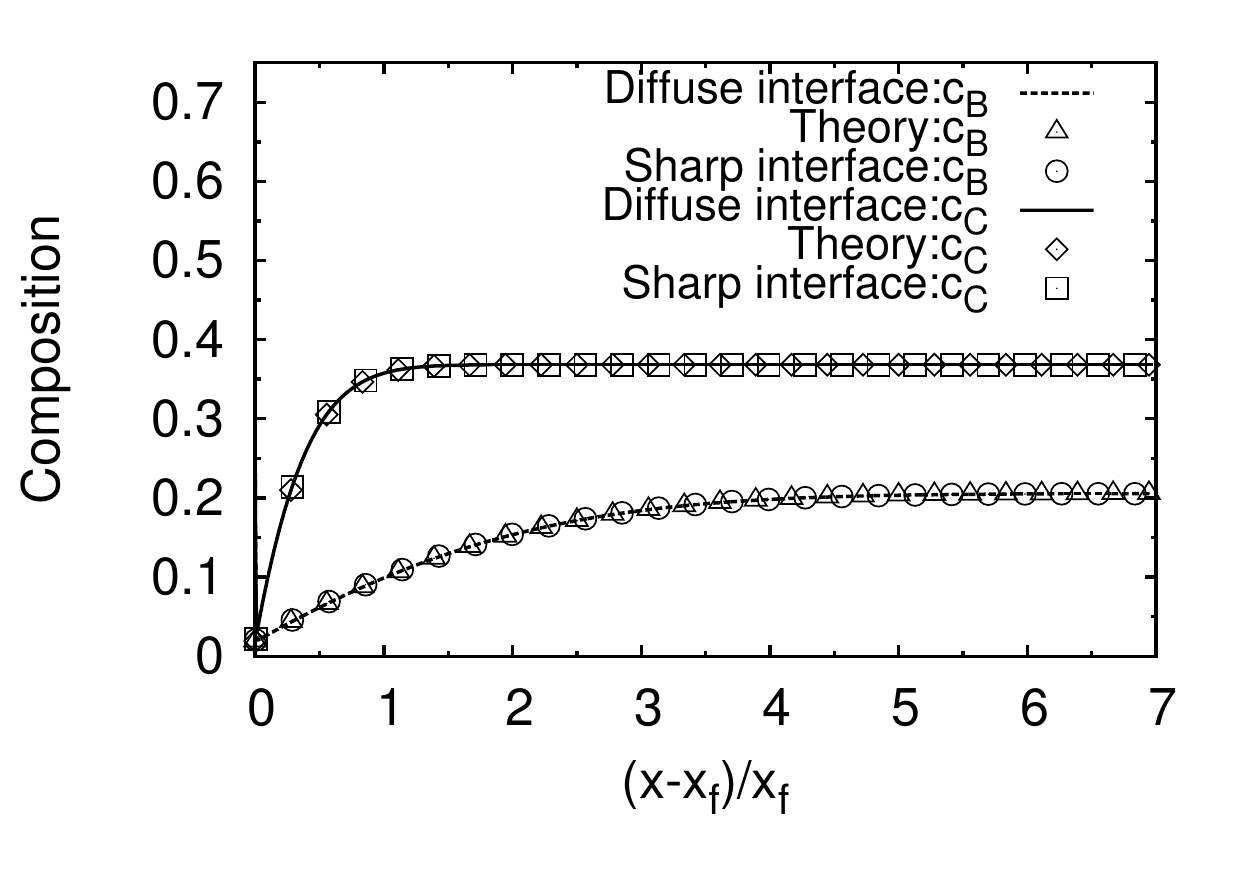}
\label{diagD_t_l_known}
}
\subfigure[]{\includegraphics[width=0.4\textwidth]{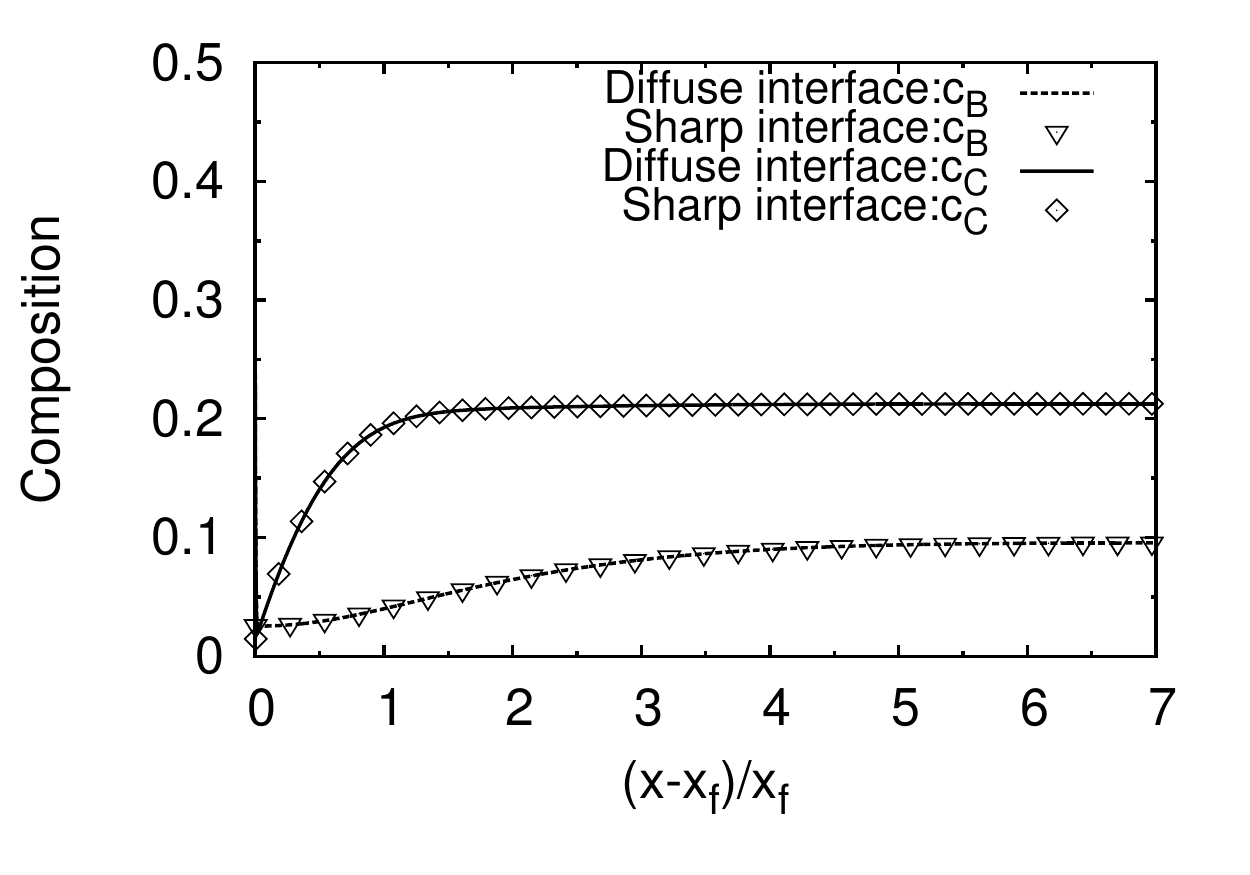}
\label{denseD_t_l_known}
}
\caption{Composition profiles at a total time of 200000 with the far-field matrix ($\alpha$) compositions 
and the diffusivity matrix components being: 
(a) $c_B=0.205$ and $c_C=0.205$; $D=I$(identity matrix),
(b) $c_B=0.205$ and $c_C=0.368$; $D_{BB}=1.0$ and~$D_{CC}=0.1$ with the off-diagonal terms set to zero, and 
(c) $c_B=0.095$ and $c_C=0.212$; $D_{BB}=1.0$ and~$D_{CC}=D_{BC}=D_{CB}=0.1$. 
The thermodynamic tie-line compositions were $c_{B,eq}^\beta=c_{C,eq}^\beta=0.481$ and 
$c_{B,eq}^\alpha=c_{C,eq}^\alpha=0.019$.  
The phase field simulation box was of size $2000$ with $dx=1.0$ and $dt=0.01$.
The $\partial c/\partial \mu$ matrix was
taken to be equal for both the phases and is stated in the Appendix red 
along with the values of $\sigma$ and $\epsilon$ that
are used in all phase field computations.}
\label{tl_known}
\end{figure}

We firstly depict a comparison between the compositions
derived from phase field computations and the theoretical predictions
for the case of the diffusivity matrix being an identity
matrix. Fig.~\ref{identD_t_l_known} highlights the excellent
agreement between theory and the phase field computations, 
where the analytical predictions are superposed on the 
phase field computations by matching the interface 
positions in the analysis and the phase field 
methods. 

As a second benchmark we choose an alloy composition 
from the theoretical analysis in Fig.~\ref{Alloy_compositions_diff_DAA_DBB}, 
which does not alter the interfacial compositions of the two phases for $D_{BB}\neq D_{CC}$.
Fig.~\ref{diagD_t_l_known}, highlights the comparison between the composition profiles
obtained from sharp interface and phase field
computations. The value of $\eta_s$ from the sharp interface ($0.625$),
phase field ($0.637$) and the theoretical analysis ($0.628$), confirm the excellent
agreement, between the three methods.


As a third benchmark, we also simulated the
influence of the presence of off-diagonal 
elements in the diffusivity matrix. The results
are depicted in Fig.~\ref{denseD_t_l_known},
which again show a good agreement between the 
sharp interface and the phase field methods.
For an arbitrary alloy composition along the 
loci of alloy compositions in Fig.~\ref{Alloy_compositions_diff_DAA_DBB},
for $D_{CC}/D_{BB}=0.1$ ($D_{CC}=0.1$), the tie-line
compositions as seen from the simulations, no 
longer remain invariant, as $c_{B,eq}^\alpha$ 
and $c_{C,eq}^\alpha$ are not equal, contrary to the property 
of the chosen thermodynamic tie-line.
Additionally, the profile
of $c_B$ shows a behavior which 
is different from the case of pure diagonal diffusivities
with a shallower increase ($\partial c_B/\partial x=0$ 
at the interface) of the compositions
near the interface. The non-existence of the composition
gradients at the interface of the 
component with a larger diffusivity is explained later.

\begin{figure}[!htbp]
\includegraphics[width=0.4\textwidth]{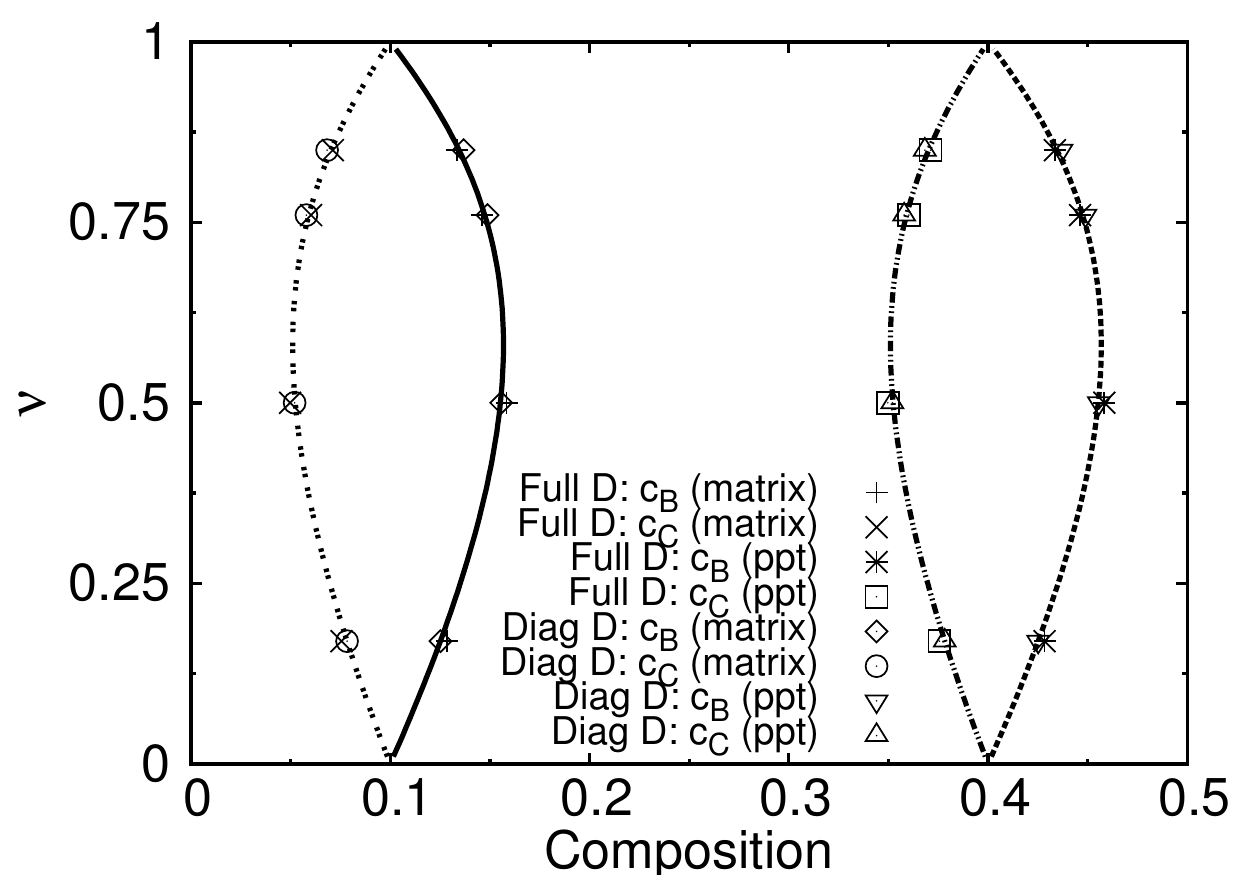}
\caption{Tie lines predicted from sharp interface simulations in 1D are compared against theoretical predictions. 
The diagonal diffusivity matrix (``Diag D'' in the figure legend) 
considered has components $D_{BB}=1.0$, $D_{CC}=0.1$ with the off-diagonal elements set to zero. 
The full diffusivity matrix (``Full D'' in the figure legend) has components  
$D_{BB}=1.0$, $D_{CC}=D_{BC}=D_{CB}=0.1$. The theoretical
predictions are depicted by the same continuous lines as done in Fig.~\ref{Tie-lines}.}
\label{planar_tie_line}
\end{figure}

Subsequently, we have chosen a far-field composition 
along a given thermodynamic tie-line, and let the simulations from 
the sharp interface and phase field methods select 
the tie-line compositions during growth. This is the inverse question, 
and more relevant to gaining a control of processing
conditions, where one can predict the precipitate
and matrix compositions given a particular alloy
composition. For the case of the diagonal diffusivities,
these tie-line compositions can also be predicted
analytically as has been previously described
in Fig.~\ref{Tie-lines}. We have superposed the 
predictions from the sharp interface simulations 
on this figure and find excellent agreement between the theoretical
and the numerical method as shown in Fig.~\ref{planar_tie_line}. 

\begin{figure}[!htbp]
\centering
\subfigure[]{\includegraphics[width=0.4\textwidth]{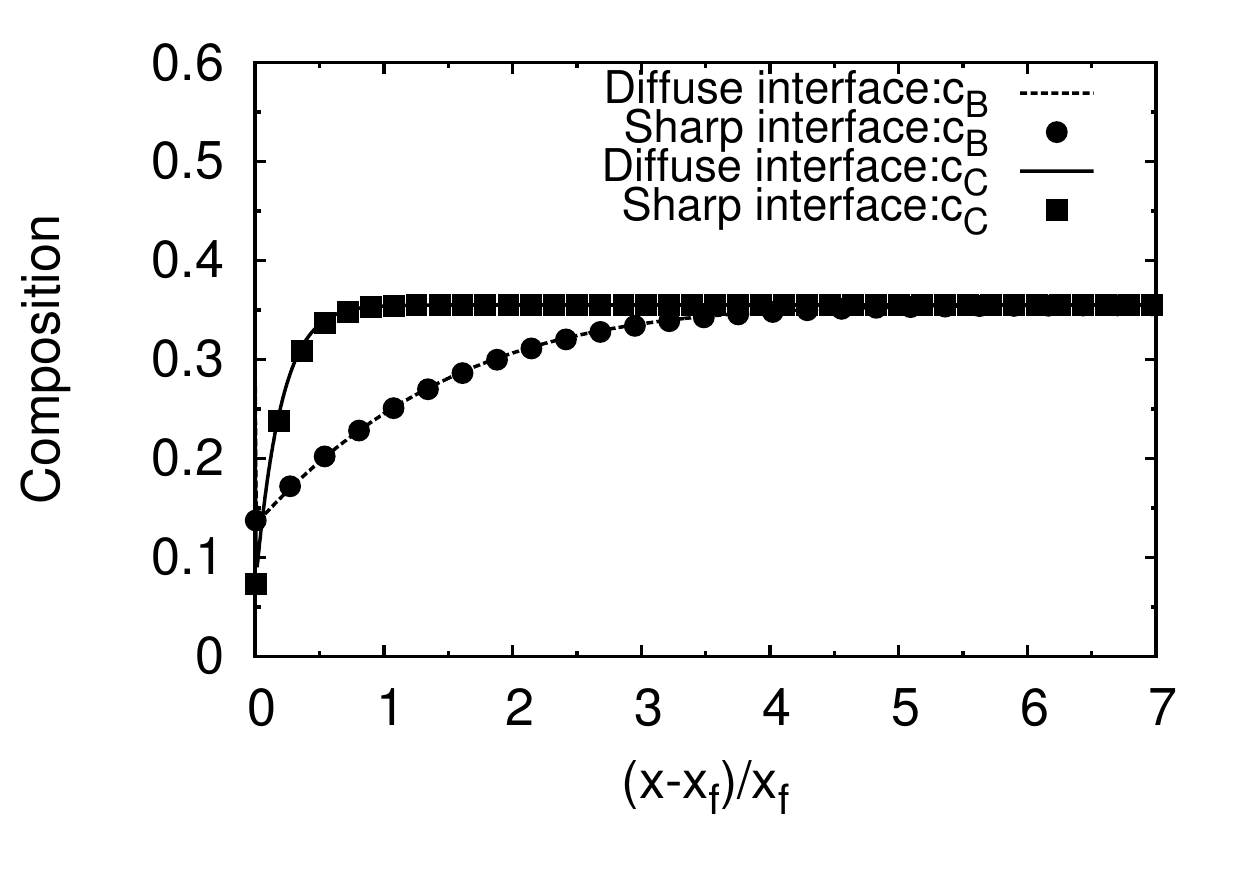}
\label{t_l_find_diagD}
}
\subfigure[]{\includegraphics[width=0.4\textwidth]{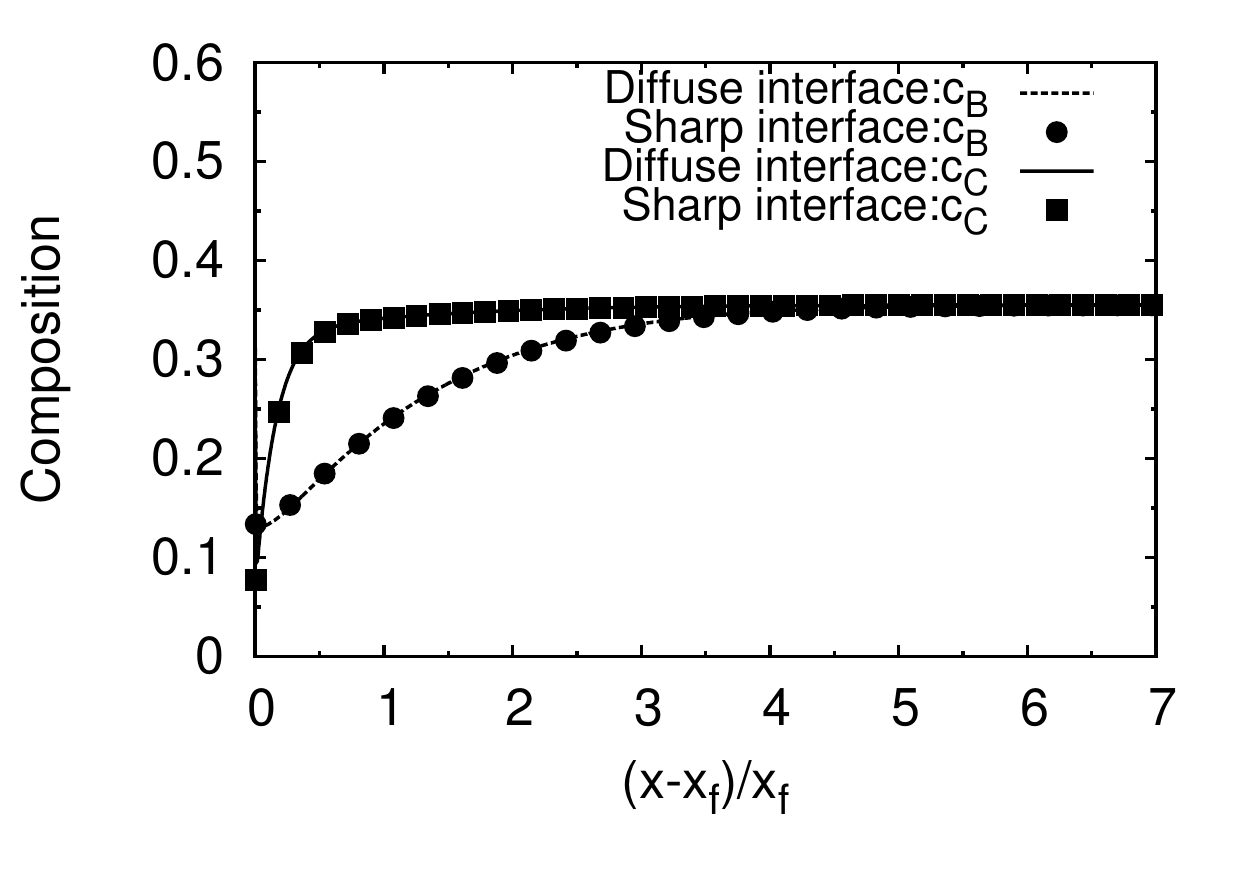}
\label{t_l_find_denseD}
}
\caption{Composition profiles at a total time of 300000 with the far-field matrix ($\alpha$) compositions 
and the diffusivity matrix components being: 
(a) $c_B=0.355$ and $c_C=0.355$; $D_{BB}=1.0$ and~$D_{CC}=0.1$ with the off-diagonal terms set to zero, and 
(b) $c_B=0.355$ and $c_C=0.355$; $D_{BB}=1.0$ and~$D_{CC}=D_{BC}=D_{CB}=0.1$. 
The other simulation parameters are the same as in Fig.~\ref{tl_known}. }
\label{tl_find}
\end{figure}

An exemplary comparison of the composition profiles
from the sharp interface and the phase field profiles, 
is depicted in Fig.~\ref{t_l_find_diagD}, for independent 
diffusion of solutes but for unequal diffusivities.

Following this, we have repeated the computations 
with off-diagonal components in the diffusivity
matrix as reported in Fig.~\ref{t_l_find_denseD}.
Here, from both the sharp interface and the 
phase field computations, we confirm that the 
selected tie-line compositions are
similar for both the pure diagonal 
diffusivities and diffusivity matrices 
with off-diagonal terms (as also seen in Fig.~\ref{planar_tie_line}).  


\subsubsection{Diffusion distances}

The diffusion length scales of the different components during growth have a profound 
impact on the kinetics of the system. For example,   
the relative influence of the components on the coarsening rates can be 
derived from the information related to the impingement of the 
composition profiles which are related to the diffusion lengths. Similarly, 
the relative sensitivity of the interface towards diffusional instabilities
can also be linked to the diffusional lengths of the components.
In a ternary system, this particular measure of the diffusion length scales 
of the different components can be 
obtained by considering the ratio of composition gradients at
the interface. This can be derived from the Stefan conditions stated in Eq.~\ref{non-normalized-stefan-condition} as,


\begin{align}
 \left\lbrace \dfrac{\partial c_i}{\partial x}\right\rbrace \Bigg|_{x_f} = -v \left[D^{-1}_{ij}\right] \left\lbrace \Delta c_j  \right\rbrace.
 \label{comp_grad}
\end{align}


From Eq.~\ref{comp_grad} we can derive 
$(\partial c_B/\partial x)/(\partial c_C/\partial x)$ 
which is the ratio of the inverses of the relative diffusion lengths
of the components $B$ and $C$. Thus, once the tie-line
compositions ($\Delta c_{B}$ and $\Delta c_{C}$) are
fixed, the ratio of the diffusion length can also 
be predicted from the preceding relation. Applying Eq.~\ref{comp_grad} on the system depicted in Fig.~\ref{t_l_find_diagD}, 
the interfacial gradients are found to be $0.000489$ and $0.00489$ in $c_B$ and $c_C$ respectively. This calculation is 
in excellent agreement to the interfacial gradients obtained from the sharp interface simulations (in phase field calculations,  
the diffuseness of the interface makes the determination of the tie-line compositions through extrapolations of
the bulk composition profiles, a little difficult). The relative diffusion 
lengths (that of $c_B$ to $c_C$), as given by the ratio of the inverses of the interfacial gradients calculates to $10$ indicating
that the component $B$ diffuses over a distance that is $10$ times over that of $C$. The same 
analysis when applied to the system in Fig.~\ref{t_l_find_denseD}, yields $\partial c_B/\partial x = 0$ and 
$\partial c_C/\partial x = 0.0054975$ at the interface (also confirmed by sharp interface calculations). 
The non-existence of an interfacial gradient in $c_B$ reflects on the dominating influence of the slower diffusing species on
the growth dynamics and that the diffusion of B is instantaneous compared to that of C and thereby
the diffusivity of C principally determines the diffusion kinetics.

\subsubsection{Effective diffusivity: Independent solute diffusion} 
A quantity which might be of interest for interpretation of the diffusion length scales
is that of the effective diffusivity ($D^{eff}_{AA}$) which is a composite diffusivity 
characterizing the $\alpha$-$\beta$ transformation. The combined diffusion of the components B and C can be mapped to a  
diffusion in the solvent A whose diffusivity 
(anointed as the effective diffusivity ($D^{eff}_{AA}$)) turns out to be a
weighted average of the diffusivities of the solutes. 
With the the diffusion of components 
B and C occurring independently, adding the 
Stefan conditions in Eq.~\ref{non-normalized-stefan-condition} gives:

\begin{align}
 v\left(\dfrac{\Delta c_B}{D_{BB}}+\dfrac{\Delta c_C}{D_{CC}} \right) = \dfrac{\partial c_A}{\partial x}\Bigg|_{x_f}= -v \dfrac{\Delta c_A}{D^{eff}_{AA}},
  \label{eff_D_1}
\end{align}
which leads to the expression:

\begin{align}
 D^{eff}_{AA}=-\left(\dfrac{1}{D_{BB}}\dfrac{\Delta c_B}{\Delta c_A} + \dfrac{1}{D_{CC}}\dfrac{\Delta c_C}{\Delta c_A}\right)^{-1}.
 \label{eff_D_2}
\end{align}

Thus, the effective diffusivity expressed in Eq.~\ref{eff_D_2} is a function of 
not only the individual inter-diffusivities of the components but also of
the chosen tie-line compositions ($c_{B,eq}^\alpha=c_{C,eq}^\alpha=0.1$ and $c_{B,eq}^\beta=c_{C,eq}^\beta=0.4$) 
As a consequence of the inverse interpolation, 
the effective diffusivity will have a value closer to the diffusivity of the slower moving species. 
Eq.~\ref{eff_D_2} predicts $D^{eff}_{AA} = 1.0$ for a system having a diagonal diffusivity 
matrix with $D_{BB}=D_{CC}=1.0$. The situation here is 
actually equivalent to the case of a binary alloy with $A$ as the diffusing species and the 
equilibrium compositions of the precipitate and the matrix being given by the tie-line 
compositions in the ternary alloy ($1-c_B-c_C$).

For a system with $D_{BB}=1.0$ and $D_{CC}=0.1$ and $D_{BC}=D_{CB}=0$ (see Fig.~\ref{t_l_find_diagD}),
$D^{eff}_{AA}$ becomes $0.182$, 
for which such a binary mapping does not hold. 
The influence of the choice of the particular tie-line on the effective 
diffusivity is also depicted in the Table~\ref{tab:d_eff}, where two 
different values of effective diffusivity are calculated based on 
two different tie-lines with the same given super-saturation. 
Please note that the $\Delta c_A$ and $\Delta c_B$ are the equilibrium 
composition values that the system chooses at the interface.

\begin{table}
\caption{Effective diffusivities of alloys having 40\% supersaturation on two different tie-lines 
(these tie-lines are used for fitting the thermodynamics). Alloys on tie-line 2 are leaner in C}
\begin{tabular}{|c|c|c|c|c|c|}
\hline 
tie-line & $\Delta c_B$ & $\Delta c_C$ & $D^{eff}_{AA}$ \\ 
\hline 
1 & -0.546 & -0.312 & 0.234 \\ 
\hline 
2 & -0.679 & 0.178 & 0.348 \\ 
\hline 
\end{tabular} 
\label{tab:d_eff}
\end{table}


\subsubsection{Effective diffusivity: Coupled solute diffusion}

For a system displaying coupled diffusion of the solutes, the gradients at the 
interface can be calculated from Eq.~\ref{comp_grad} and ($D^{eff}_{AA}$) can be computed as:

\begin{align}
 \dfrac{\partial c_B}{\partial x}\Bigg|_{x_f}+  \dfrac{\partial c_C}{\partial x}\Bigg|_{x_f}=
 -\dfrac{\partial c_A}{\partial x} \Bigg|_{x_f} = v \dfrac{\Delta c_A}{D_{eff}^{AA}},
 \label{effectD}
\end{align}
which leads to:
\begin{align}
 D_{AA}^{eff}=\dfrac{v \Delta c_A}{\dfrac{\partial c_B}{\partial x}\Bigg|_{x_f} +  \dfrac{\partial c_C}{\partial x}\Bigg|_{x_f}} =  \nonumber \\
 \dfrac{- \Delta c_A \left(D_{BB}D_{CC}-D_{BC}^2\right)}{\left(D_{CC}-D_{BC}\right)\Delta c_B + \left(D_{BB}-D_{BC}\right)\Delta c_C},
  \label{effectD2}
\end{align}

where we have imposed $D_{BC}=D_{CB}$. 
For a system with $D_{BB}=1.0$, $D_{CC}=0.1$, $D_{BC}=D_{CB}=0.1$ (see Fig.~\ref{t_l_find_denseD}), 
the effective diffusivity was calculated to be $0.2$.

Thus, through our study in 1D, we have been able to predict the new equilibrium phase compositions selected 
during growth. 
Furthermore, strategies for computing the solute diffusion distances in the matrix are also discussed. Though,
this information is critical to an understanding of the microstructural length scale selection during growth, it must be 
complemented by an analysis of the role played by capillarity in dimensions greater than one. 
This leads us to take up the same problem again, but in 2D.

\section{Radial growth of a cylindrical precipitate}
The subject of interest is dealt in a manner similar to planar growth study. 
An analytical theory is developed followed by a description of the 
sharp interface technique in 2D. The phase field model has already been 
explained in conjunction with the 1D problem and is not discussed
here. The description of the system thermodynamics is modified to 
account for the effect of curvature.  

\subsection{Theory}
In the radial coordinate system, the governing differential equations are written in the vector-matrix form as,
\begin{align}
 \left\lbrace\dfrac{\partial c_i}{\partial t}\right\rbrace &=  \dfrac{1}{r} \left[D_{ij}\right] \left\lbrace \dfrac{\partial}{\partial r} \left( r \dfrac{\partial c_j}{\partial r}\right)\right\rbrace.
 \label{non-normalized-governing-equations_radial}
\end{align}
The Stefan boundary condition 
at the interface ($r=R$) in the radial coordinate system is the same 
as in Eq.~\ref{non-normalized-stefan-condition} with $r$ in place of $x$. 


Restricting ourselves to diagonal diffusivities, a co-ordinate transformation 
of Eq.~\ref{non-normalized-governing-equations_radial} to express them as functions of  
$\eta = r/\sqrt{t}$ followed by an integration with respect to $\eta$ leads to,

\begin{align}
 \left\lbrace\dfrac{\partial c_i}{\partial \eta} \right\rbrace&= \left\lbrace\dfrac{\lambda_i^{R}}{\eta} \exp\left(\dfrac{-\eta^{2}}{4D_{ii}}\right)\right\rbrace,
 \label{integrate_once_radial}
\end{align}





where $\lambda_i^{R}$'s are integration constants. Using the Stefan's conditions in 
Eq.~\ref{Stefan_condition}, the value of the integration constants can 
be derived as, 

\begin{align}
 \left\lbrace\lambda_i^{R} \right\rbrace &= \dfrac{-\eta_s^2 }{2} \left\lbrace\dfrac{\Delta c_{i} \left( R \right)}{D_{ii} \exp\left(\dfrac{-\eta_s^{2}}{4 D_{ii}}\right)}\right\rbrace,
 \label{rad_const}
\end{align}

where $\eta_s = R/\sqrt{t}$ is the corresponding value at the interface, 
which is at a position $R$ at a given time $t$ and the definition of 
$\Delta c_i$ remain the same as in 1D, except that 
now they are functions also of the radius of the precipitate through
the Gibbs-Thomson effect, i.e the compositions $c_{i,eq}^{\alpha,\beta}$
are functions of the radius of the precipitate.

Far into the growth regime, the composition differences $\Delta c_i$,
vary very slowly upon change of radius, therefore can be treated as constants 
and since $\lambda_i^R$ are constants independent of $\eta$ 
the only possibility is that $\eta_s$ be a constant, 
for such a scaling regime to exist.

Integrating Eq.~\ref{integrate_once_radial} from $\eta_s$ to $\infty$, 
we can derive using Eq.~\ref{rad_const}, 

\begin{align}
 \left\lbrace\dfrac{c_i^{\infty} - c_{i,eq}^{\alpha}\left(R\right)}{\int_{\eta_s}^{\infty}\dfrac{1}{\eta}\exp\left(-\dfrac{\eta^{2}}{4D_{ii}}\right)d\eta}\right\rbrace 
 &= \dfrac{-\eta_s^2}{2}\left\lbrace\dfrac{ \Delta c_{i}\left(R\right)}{D_{ii} \exp\left(\dfrac{-\eta_s^{2}}{4D_{ii}}\right)}\right\rbrace, 
 \label{composition_profiles_radial}
\end{align}

Particularizing Eq.~\ref{composition_profiles_radial} to ternary systems, 
we can see that while in the 1D case, employing the functions $c_B^{\alpha,\beta}(\mu_B,\mu_C)$, $c_C^{\alpha,\beta}(\mu_B,\mu_C)$ 
and $\mu_{B,eq}(\mu_{C,eq})$, yields a system of non-linear equations which can be solved for 
$\mu_C$ and $\eta_s$, leading to the equilibrium tie-line compositions,
however, for the case of the cylindrical precipitate this is no longer possible.
This is because, though the system selects a particular 
value of $\eta_s$ during the scaling regime of 
precipitate growth, it can correspond to a larger $R$ at a later time or a smaller $R$ at an earlier time. 
This makes it impossible to determine the value of the equilibrium
compositions at the interface without the knowledge of the 
radius $R$. 

Despite this constraint, we can still attempt to understand the 
effect of dimensionality by ignoring the curvature effect on compositions. 
Under this assumption, Eq.~\ref{composition_profiles_radial} are solved for $\mu_C$ and $\eta_s$,
leading to curves in Fig. \ref{etas_vs_nu}, where the growth coefficient
is derived for compositions along a given thermodynamic
tie-line with diagonal diffusivity matrices described in the Fig. \ref{etas_vs_nu}.
Correspondingly, one can also predict the selected tie-line compositions similar 
to the computations for case of planar growth as in Fig.\ref{Tie-lines}.

However, if the equilibrium compositions at the interface 
$c_{B,eq}^{\alpha}, c_{C,eq}^{\alpha}$  
are known, then Eq.~\ref{composition_profiles_radial} can be utilized for generating the 
composition profiles in the matrix phase $\alpha$. 
At any particular instant of time, all $\eta$'s which are $>\eta_s$ 
can be mapped to locations ahead of the interface (i.e., inside the matrix by using $r=\eta \sqrt{t}$) 
with $c_B(\eta)$ and $c_C(\eta)$ being the compositions at those locations. Integrating 
Eq.~\ref{integrate_once_radial} from any particular $\eta$($>\eta_s$) to $\infty$, 
we can derive using Eq.~\ref{rad_const}, 



\begin{align}
\left\lbrace\dfrac{c_i^{\infty} - c_{i,eq}^{\alpha}\left(R\right)}{c_i^{\infty} - c_i\left(\eta\right)}\right\rbrace=
\left\lbrace Q_i\right\rbrace=
\left\lbrace\dfrac{\int_{\eta_s}^{\infty}\dfrac{1}{\eta}\exp\left(-\dfrac{\eta^{2}}{4D_{ii}}\right)d\eta}
{\int_{\eta}^{\infty}\dfrac{1}{\eta}\exp\left(-\dfrac{\eta^{2}}{4D_{ii}}\right)d\eta}\right\rbrace, 
\label{rad_ratio}
\end{align}

where $Q_i$'s are constants representing ratios of integrals. 
Eq.~\ref{rad_ratio} can be re-written to obtain $c_B(\eta)$ and $c_C(\eta)$:

\begin{align}
 \left\lbrace c_i(\eta)\right\rbrace
 =\left\lbrace\dfrac{c_{i,eq}^{\alpha}\left(R\right)}{Q_i}+\left(1-\dfrac{1}{Q_i}\right)c_i(\infty)\right\rbrace, \nonumber \\
 \label{rad_prof}
\end{align}

and this can be done for all $\eta>\eta_s$ to get the composition profiles in the matrix. 

\subsection{Sharp-interface model}
Eq.~\ref{non-normalized-governing-equations_radial} can be numerically solved 
for by discretizing them in a frame attached to the 
interface as was done for the 1D case. The governing equations can 
be re-written in the matrix-vector notation for a moving frame of reference as:
 

\begin{equation}
 \left\lbrace\dfrac{\partial c_i}{\partial t} -v\dfrac{\partial c_i}{\partial r}\right\rbrace = \left[D_{ij}\right]
 \left\lbrace\left(\dfrac{1}{r}\dfrac{\partial c_j}{\partial r}+\dfrac{\partial^2 c_j}{\partial r^2}\right)\right\rbrace,
 \label{rad_sh}
\end{equation}
where $v$ is the velocity of the interface at a particular instant of time.
Similar to the situation in 1D, this requires the determination of 
the interfacial compositions at each time step, which are solved 
by utilizing the Stefan conditions along with the conditions for 
local thermodynamic equilibrium. 
In contrast to the situation in 1D where thermodynamic
equilibrium is derived by setting the driving force to zero, 
in 2D, the same is derived by 
equating the driving force due to phase transformation with that due to 
curvature, i.e., 
$\Delta \Psi^{\alpha\beta}=\sigma\kappa$, that reads,
\begin{align}
 \dfrac{1}{V_m}\left\lbrace c_{i,eq}^{\beta,*} - c_{i,eq}^{\alpha,*}\right\rbrace \left\lbrace\mu_i-\mu_{i,eq}^{*}\right\rbrace = \sigma\kappa,
 \label{rad_equi}
\end{align}

where $\sigma$ denotes the interfacial energy and $\kappa$ the curvature which can 
be approximated by $1/R$ for a cylindrical precipitate. The
expression in Eq.~\ref{rad_equi} can be re-written for a ternary
alloy to obtain an expression relating the two diffusion potentials 
$\mu_B$ and $\mu_C$ for the case of two independent components 
as,

\begin{align}
 \mu_B-\mu_{B,eq}^{*}=\dfrac{\sigma \kappa V_m}{(c_{B,eq}^{\beta,*}-c_{B,eq}^{\alpha,*})} 
 - \dfrac{(c_{C,eq}^{\beta,*}-c_{C,eq}^{\alpha,*})(\mu_C-\mu_{C,eq}^{*})}{(c_{B,eq}^{\beta,*}-c_{B,eq}^{\alpha,*})}.
 \label{rad_mu_A}
\end{align}

Using this relation and the relations $c_i^{\alpha}\left(\vmu\right)$ one can reduce the 
Stefan boundary conditions at the interface purely as functions of one of the diffusion 
potentials and solve for $\mu_C$ and the velocity in a manner similar to the situation in 1D
\footnote{$\Delta c_B$ and $\Delta c_C$ can vary with $R$ during growth when the $\partial c /\partial \mu$ matrix is different for the 
matrix($\alpha$) and the precipitate ($\beta$) phases. For such a system, the scaling given by the constancy of of $\lambda_B^{R}$ 
and $\lambda_C^{R}$ in Eq.~\ref{rad_const} happens later in time, characterized by a slow
change in $c_B(\eta_s)$ and $c_C(\eta_s)$ resulting in an asymptotic 
approach to a constant $\Delta c_B$ and $\Delta c_C$.}.


\subsection{Results}
\begin{figure}[]
\centering
\subfigure[]{\includegraphics[width=0.4\textwidth]{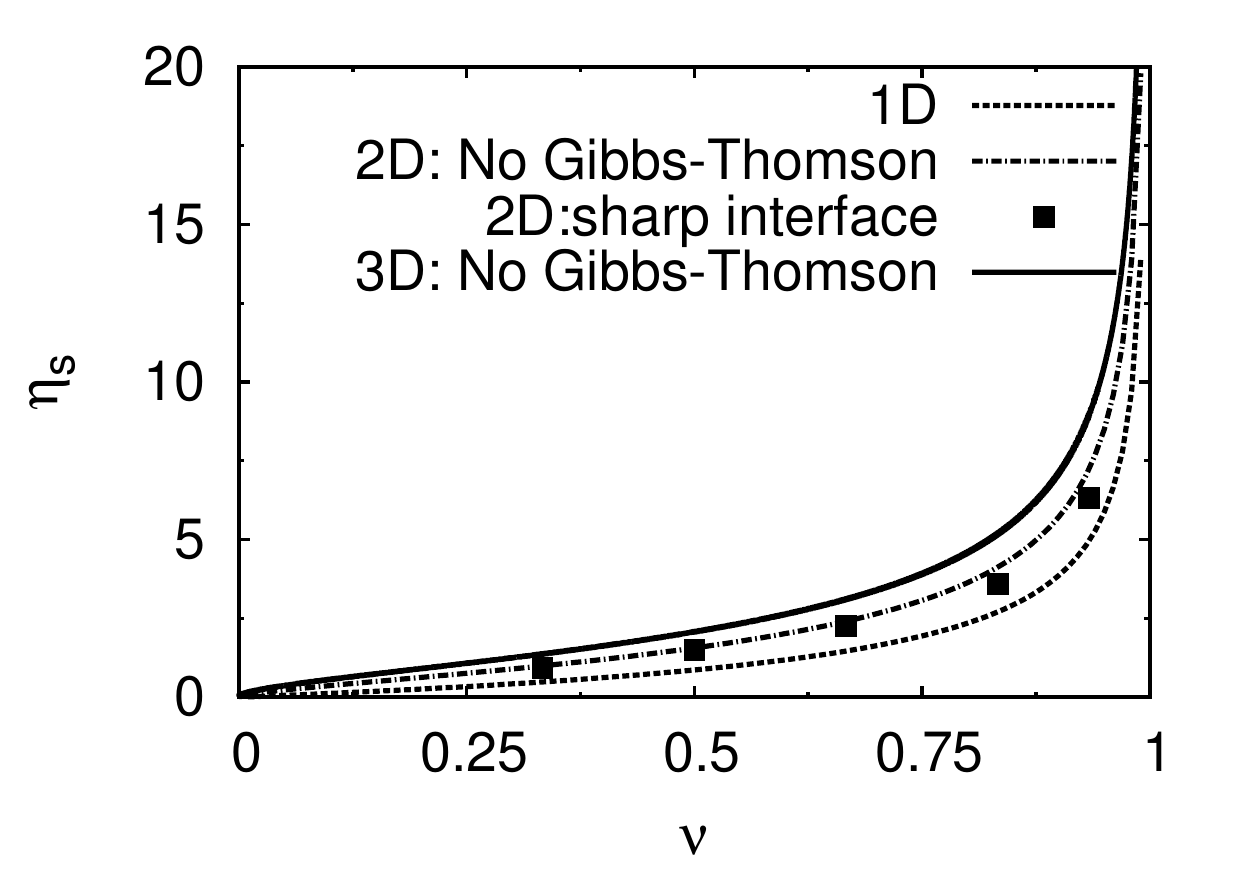}
\label{eta_vs_nu_deqi}
}
\subfigure[]{\includegraphics[width=0.4\textwidth]{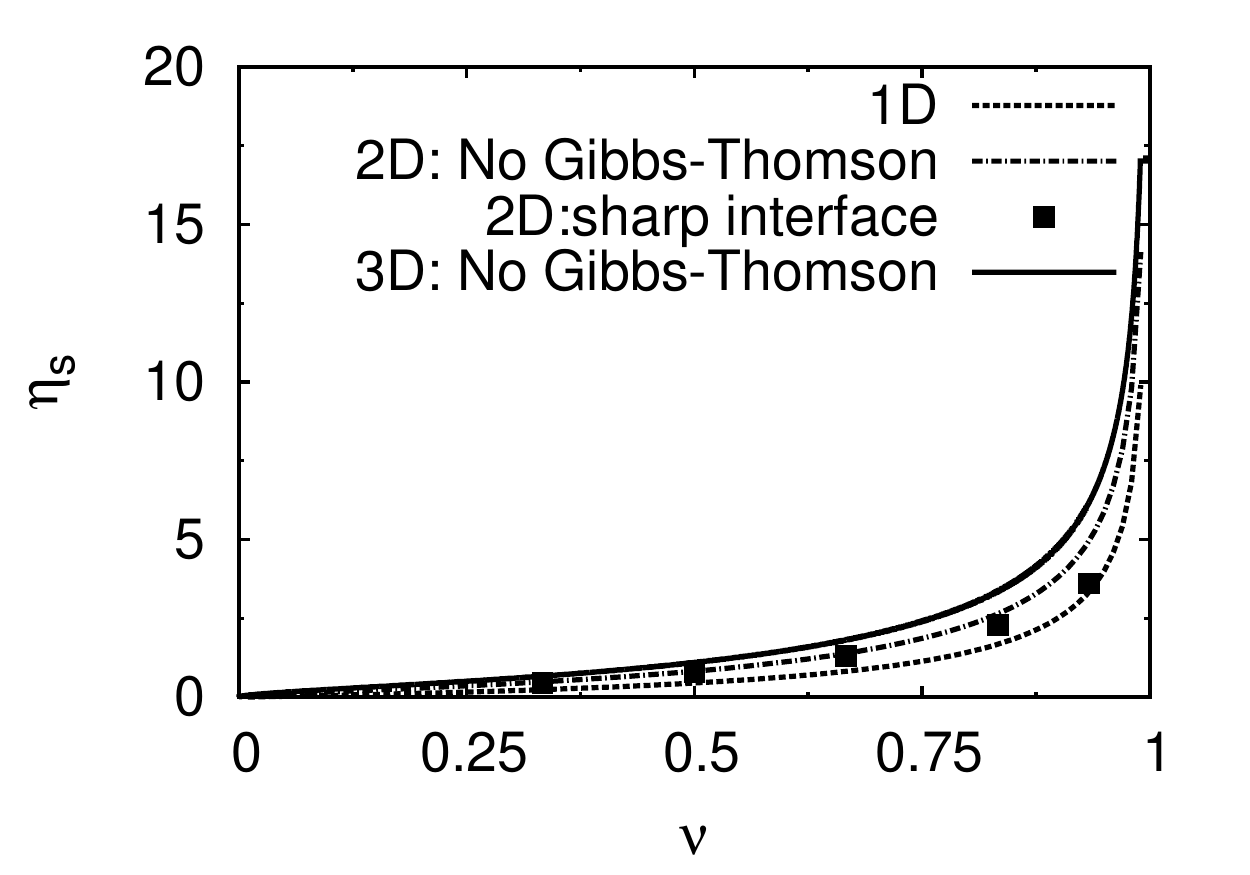}
\label{eta_vs_nu_dratio01}
}
\caption{Variation in $\eta_s$ with $\nu$, for a 2D system with Gibbs-Thomson correction from sharp interface calculations,
for (a) $D_{BB}=1.0$ and $D_{CC}=1.0$, and (b) $D_{BB}=1.0$ and $D_{CC}=0.1$. 
The other terms in the diffusivity matrix are zero.
The results are presented in the context of similar variations 
obtained for 1D as well as for 2D and 3D systems neglecting the influence of 
capillarity. }
\label{etas_vs_nu}
\end{figure}

The implication of considering the correction in the local equilibrium due to curvature 
can be understood by first solving for $\eta_s$
from Eq.~\ref{composition_profiles_radial} ignoring the influence of curvature 
on the shift of compositions, for different supersaturations defined by $\nu$, 
super-imposed with points computed from values obtained from sharp interface
computations in 2D incorporating capillarity. 
The variation in $\eta_s$ against $\nu$ (whose definition is described in the caption to Fig.~\ref{Tie-lines}) 
thus calculated is studied in the context of similar variations obtained from 
solving the 1D problem (see Fig.~\ref{etas_vs_nu}).
Though, we do not address the 3D problem in this study, a variation of $\eta_s$ 
with $\nu$ for a growth of a spherical particle 
is also presented in Fig.~\ref{etas_vs_nu} for the sake
of completion. 

We see that differences between the analytical predictions
without consideration of capillarity and sharp interface computations
including capillarity occur for the case where $D_{BB}\neq D_{CC}$ at
large volume fractions, while for smaller values of $\nu$, the
deviations are small. This implies, that the selection of 
the growth coefficient $\eta_s$ is only weakly influenced by capillarity.

\begin{figure}[]
\centering
\subfigure[]{\includegraphics[width=0.4\textwidth]{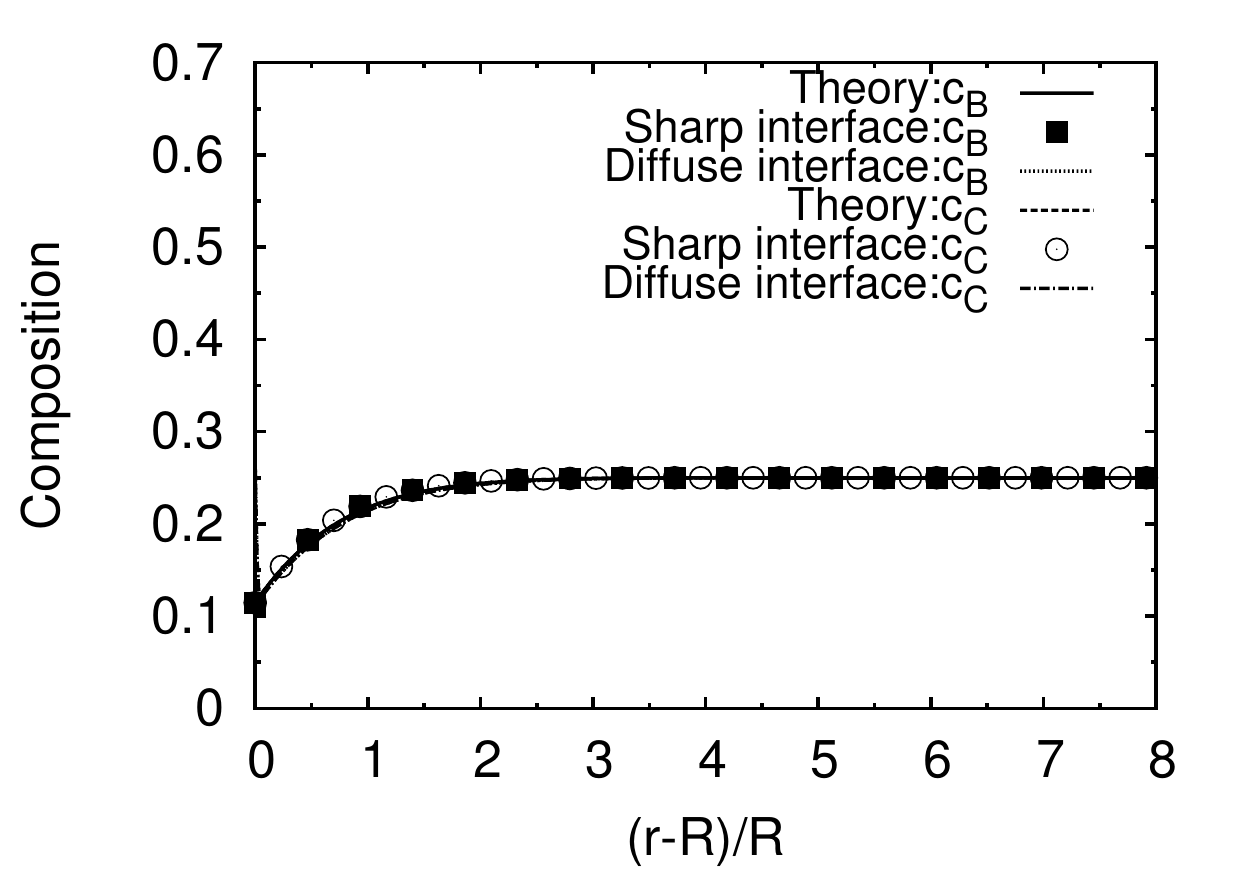}
\label{radial_D_eq_I}
}
\subfigure[]{\includegraphics[width=0.4\textwidth]{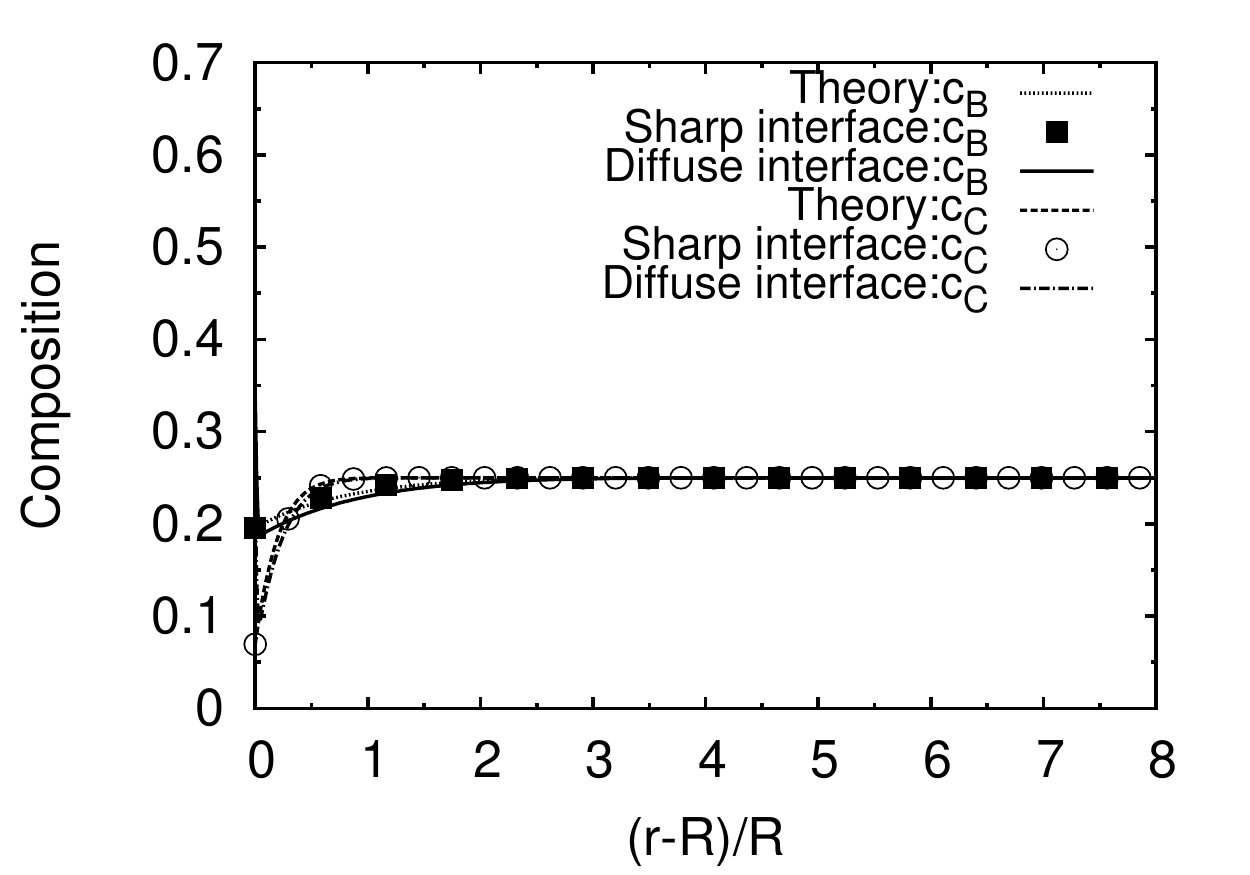}
\label{radial_Dbb_10pc}
}
\caption{Composition profiles at a t=10000 with the far-field liquid compositions being $c_B=0.25$ and $c_C=0.25$. 
The diffusivity matrix is the
same as an identity matrix in (a), while it is $D_{BB}=1.0$ and $D_{CC}=0.1$ with the off-diagonal entries zero for (b). 
The simulation is performed on an $800 \times 800$ box 
with $dx=dy=1.0$, $dt=0.01$, with the same maintained for sharp interface calculations as well. }
\label{radial_comp}
\end{figure}

\begin{figure}[!htbp]
\centering
\subfigure[]{\includegraphics[width=0.4\textwidth]{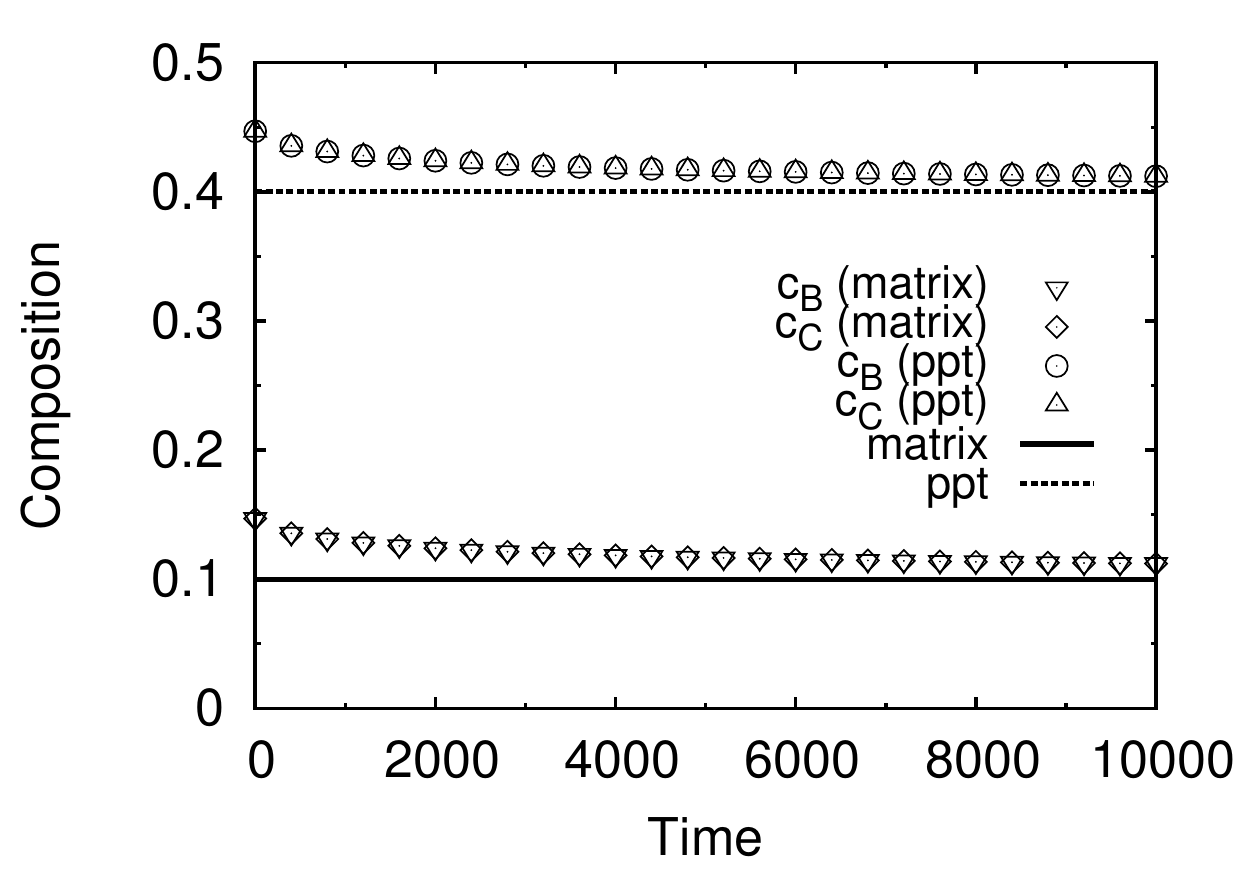}
\label{2d_cB}
}
\subfigure[]{\includegraphics[width=0.4\textwidth]{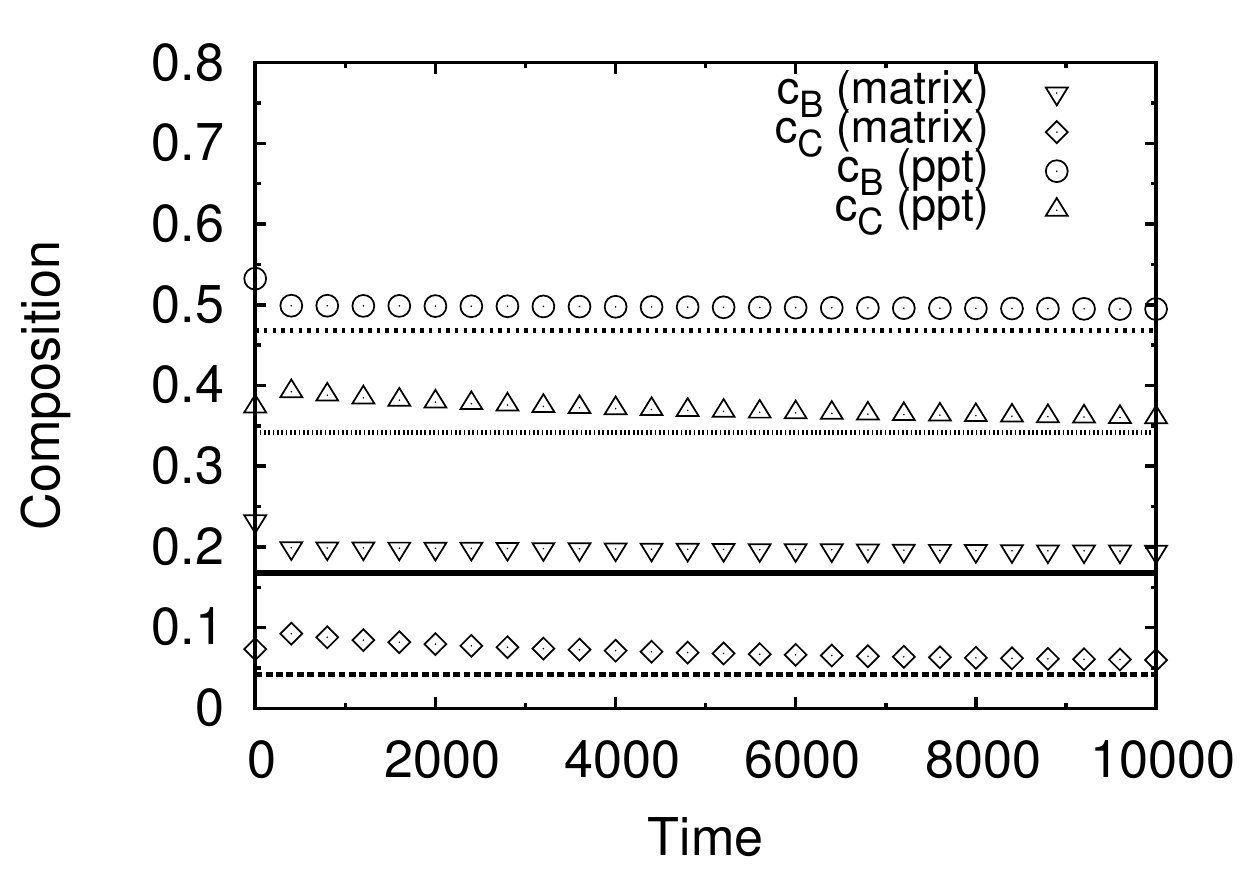}
\label{2d_cC}
}
\caption{Variation of tie-line compositions with time, with the non-zero 
diffusivity components being, (a)$D_{BB}=D_{CC}=1$, and (b) $D_{BB}=1$, $D_{CC}=D_{BC}=D_{CB}=0.1$,
during radial growth of a precipitate. The lines plotted along side the data-points refer
to the analytical calculations without incorporation of capillarity. (time is plotted 
in non-dimensional units)}
\label{radial_tie_lines}
\end{figure}

Further, phase field and sharp interface numerical simulations both incorporating
the influence of capillarity are utilized for deriving the composition profiles 
far into the scaling regime. These profiles are compared with analytical 
predictions obtained by solving Eq.~\ref{rad_prof}, where we set the values of 
equilibrium compositions at the interface, 
$c_{B,eq}^{\alpha}$ and $c_{C,eq}^{\alpha}$ and $\eta_s$ from sharp interface calculations.
Fig. \ref{radial_tie_lines} shows the transient evolution of the interfacial 
compositions obtained from the numerical sharp interface simulations.
In Fig.~\ref{radial_D_eq_I}, the diffusivity matrix is diagonal and the individual
diffusivities are set to unity. 
Here, the composition profiles from the three techniques mentioned above are in excellent agreement with each other. The values
of $\eta_s$ selected by the system is $1.42$($1.51$) as obtained from phase field (sharp interface) simulations.
The changing interface curvature of a growing precipitate sets the interfacial 
compositions (under local thermodynamic equilibrium) 
along an extension to the original tie-line only and 
the system does not select a tie-line with a different $c_B/c_C$ ratio 
(see Fig.~\ref{2d_cB}, where $c_B=c_C$ is the original tie-line).

Focussing on the composition profiles for the case of $D_{BB}/D_{CC}=10$ ($D_{CC}=0.1$), we can see from 
Fig.~\ref{radial_Dbb_10pc}, that the interfacial compositions
in the matrix do not correspond to the original tie-line where $c_B/c_C=1.0$. 
This can also be observed from both the analytically calculated
curves (neglecting Gibbs-Thomson) and the data points obtained from sharp interface calculations, for a
particular value of $\nu$ as seen in Fig.\ref{2d_cC}.

In addition, as expected, the deviations, of tie-line compositions obtained using numerical simulations
(considering capillarity), from analytical predictions without incorporating 
capillarity, reduce with time as the ratio of 
the radius of the precipitate with respect to the capillary 
length (approximately scaling as: $\left(\dfrac{\sigma V_m}{\left(d\mu/dc\right)\Delta c}\right)$) 
becomes larger as seen in Fig.\ref{radial_tie_lines}.

The lowered gradients in $c_B$ in Fig.~\ref{radial_Dbb_10pc} at the interface translates to a 
diffusion distance large enough to interact with the system boundaries.
This can explain the slight difference between the sharp interface (also theoretical) 
and the phase field profiles of component B. More elaborately, the differences
arise because it is difficult to impose equivalent boundary conditions between a
radial co-ordinate system that is used for both the sharp interface and the 
theoretical calculations (which agree well) and a cartesian co-ordinate system 
in a rectangular domain that is used for the phase field. This difference
causes a small error for the diffusion profiles
with shallower gradients, possibly due to the different interaction with 
the boundaries in the two co-ordinate systems.
The $c_C$ profiles (displaying larger gradients at the interface)
obtained from the different schemes described above are in very good agreement 
with each other as can be seen from Fig.~\ref{radial_Dbb_10pc}.      
Both numerical schemes (phase field and sharp interface) 
predicted the same value of $\eta_s=0.72$ for this system.
The improved match in the $\eta_s$'s from the 
phase field and the sharp interface calculations in this situation compared to the one 
where $D_{BB}/D_{CC}=1$ can be attributed to the 
fact that in the former, the slower diffusing species controls 
the growth of the precipitate (which happens at
a much slower rate than in the case where $D_{CC}=1$). 

Taking cue from the minor changes observed 
in tie-line selection due to the presence
of off-diagonal terms in the diffusivity matrix in 1D, we restrict 
our studies in 2D to diagonal diffusivity matrices only, knowing that the sharp
interface and phase field simulations can be easily extended to capture 
the dynamics corresponding to a diffusivity matrix with off-diagonal entries.

\section{Summary and Discussion}
In summary, we use numerical simulation methods 
(phase field, sharp interface) and analytical calculations
for the determination of phase equilibria in multi-component
systems. Here, firstly we give a prescription for the growth of
planar interfaces, where starting from the thermodynamics of the system 
in terms of the free-energies near the co-existence compositions it is possible to 
analytically derive not only the tie-line compositions that the phases 
will select for a given ratio of diffusivities in a diagonal
diffusivity matrix but also the range of alloy compositions sharing the 
same given tie-lines. These predictions agree well against numerical 
phase field and sharp interface calculations. Additionally, numerical 
computations have been utilized to extend the study to include 
the case of full diffusivity matrices which are difficult to 
treat analytically. 

Thereafter, we investigate the growth of cylindrical precipitates in 2D, 
where our analytical and sharp-interface calculations allow the 
equilibrium compositions of the phases to vary with curvature.
This allows us to capture the continuous selection of different tie-lines 
during growth of precipitates for different diffusivity ratios. 
Here as well, the composition profiles obtained from our theory, sharp interface and 
phase field calculations agree well with each other at any given instant of time.
The influence of the incorporation of the Gibbs-Thomson can be seen in 
Fig. \ref{eta_vs_nu_dratio01}, where the deviations of the growth coefficient
obtained from sharp interface simulations, from the predictions without the consideration of the Gibbs-Thomson 
corrections occur at very high volume fractions. Consequently, we can 
derive that the differences in the predictions of the phase equilibria 
(with and without Gibbs-Thomson correction in 2D) seem
to influence the growth coefficient $\eta_s$ only weakly at low volume 
fractions.

The diffusion distances in the matrix for different solutes 
are also computed which are critical to the prediction of the onset of coarsening. 
Considering the large magnitude of the diffusion length scales calculated in this paper, 
they appear to be a quantity easily obscured in experimental studies of multi-particle precipitation
where the inter-particle distance is not large enough to resolve the steady-state growth regime from
coarsening. Thus, unless experiments are designed specifically to measure diffusion distances during growth, 
analytical and numerical techniques discussed in this paper provide the only methodologies for 
computing this important parameter. Furthermore, an effective diffusivity, 
defined as an average of the individual solute diffusivities weighted 
by the tie-line compositions is used to represent the overall 
kinetics of the system.

It is important to highlight at this point 
that the $\partial c/\partial \mu$ matrix plays an important role in the selection 
of tie-lines with different $c_B/c_C$ ratios with time. For a precipitate phase 
which is an intermetallic (very limited solubility around its stoichiometric composition), 
the components of the $\partial c/\partial \mu$ matrix
are expected to be very small in magnitude, resulting in negligible migration of the tie-lines
to other $c_B/c_C$ ratios during growth.
Analytical and numerical techniques proposed in this paper
can help delineate the bulk alloy compositions that ensures growth of such phases with the desired composition.
This information can facilitate a stringent control during processing to achieve the microstructural objectives.
Additionally, it is noteworthy that the simple
scheme of incorporating the thermodynamics for (both phase field
and sharp interface methods) is similar to 
previous work \cite{Choudhury+15} where parabolic 
free-energy extrapolations are used, while for the present
paper, we restrict ourselves to linearized driving forces. 
The simplicity of the framework allows for easy extension to multicomponent
alloys for more than three components and incorporation of information from 
thermodynamic databases. However, the accuracy of the 
assumption of linearization, must be checked depending on the deviations
of phase equilibria from those around which the linearization is
performed.

\section{Conclusion and outlook}
To conclude we have the following inferences from the work

\begin{itemize}
\item A particular phase field model based on a grand-potential formulation is validated for
simulating phase transformations in multicomponent systems, thus setting up the modeling
of more complicated kinetic processes such as coarsening and growth in multi-phase
systems which are beyond the purview of sharp interface and analytical considerations.

\item Analytical approaches for diagonal diffusivity matrices are outlined which allow
for the determination of the phase equilibria using the same thermodynamic information
also utilized in the phase field simulations, i.e., the information relating to the 
properties of the free-energies of the respective phases with composition.

\item A strong implication of the work is the need for accurate measurements/determination
of diffusivity/mobility matrices without which results from numerical simulations
become less useful in the quantitative understanding of growth in multi-component
systems.
\end{itemize}

A corollary that can be derived of the present work is highlighting the important difference
between binary alloys and ternary (and higher) systems: Most growth relations
for interfaces (interface response functions) relate the velocity of the interface for different morphologies
such as lamellar, dendritic etc, with the imposed thermodynamic
conditions such as undercooling or supersaturation. These relations which have
been derived for binary alloys will have to be modified to include the extra degree of freedom 
that allows for the choice of equilibrium compositions to depend on the diffusivity matrices
for systems with greater than two components. This presents an exciting scope for future work.

\section*{Acknowledgements}
The authors thank Department of Science and Technology for support through
its Thematic Unit of Excellence program in computational materials
science. 

\section{Appendix}

\subsection{Thermodynamic information}
The $\partial c/\partial \mu$ matrix was kept constant over both the phases and this is a thermodynamic parameter 
that was common to all calculations: 

\begin{equation}
\dfrac{\partial c}{\partial \mu} = 
\left[ \begin{array}{cc}
1.1548 & 0.0535  \\
0.0535 & 1.0025  \\
\end{array} \right]
\end{equation}

\subsection{Interfacial energy and width in phase field simulations}
The interfacial energy and width are determined in all these simulations by using $\sigma=1$ and $\epsilon=4$ (which corresponds 
to about ten points in the interface for a grid resolution dx=1).

\subsection{Non-dimensionalization}
All values reported in this paper are non-dimensionalized. 
By choosing appropriate length, time and energy scales characterizing a particular
system, we can retrieve dimensional quantities describing the growth behaviour of that system.
The length $l^{*}$, time $t^{*}$ and energy scales $f^{*}$ are defined as,
\begin{align}
 f^* &= \dfrac{1}{V_m}\left[\dfrac{\partial \mu_i}{\partial c_j}\right]_{max}, \\
 l^* &= \dfrac{\sigma}{f^*}, \\
 t^* &= \dfrac{{l^*}^2}{\left[D_{ij}\right]_{max}}.
 \label{non-dimensionalization}
\end{align}

\bibliography{references2_mod}

\end{document}